%% file: EXO-17-001_temp.tex
\pdfoutput=1

\documentclass[11pt,twoside,a4paper,cmspaper,final,collab]{cms-tdr}
\def\svnVersion{443428P}\def\cmsCernNoTag{CERN-EP-2017-235}\def\cmsCernDate{\today}\def\cmsMessage{Published in the Journal of High Energy Physics as \href{http://dx.doi.org/10.1007/JHEP01(2018)097}{\doi{10.1007/JHEP01(2018)097.}}}

\begin{document}\cmsNoteHeader{EXO-17-001}

\hyphenation{had-ron-i-za-tion}
\hyphenation{cal-or-i-me-ter}
\hyphenation{de-vices}

\RCS$Revision: 437287 $
\RCS$HeadURL: svn+ssh://svn.cern.ch/reps/tdr2/papers/EXO-17-001/trunk/EXO-17-001.tex $
\RCS$Id: EXO-17-001.tex 437287 2017-12-02 18:19:41Z cmantill $
\newlength\cmsFigWidth
\ifthenelse{\boolean{cms@external}}{\setlength\cmsFigWidth{0.85\columnwidth}}{\setlength\cmsFigWidth{0.4\textwidth}}
\ifthenelse{\boolean{cms@external}}{\providecommand{\cmsLeft}{top\xspace}}{\providecommand{\cmsLeft}{left\xspace}}
\ifthenelse{\boolean{cms@external}}{\providecommand{\cmsRight}{bottom\xspace}}{\providecommand{\cmsRight}{right\xspace}}
\providecommand{\NA}{{\ensuremath{\text{---}}}}
\newcommand{\cmsptmiss}{\ensuremath{p_{\mathrm{T}}^{\mathrm{miss}}}\xspace}
\newcommand{\x}{\ensuremath{\phantom{0}}}
\newcommand{\y}{\ensuremath{\phantom{.}}}

\cmsNoteHeader{EXO-17-001}
\title{Search for low mass vector resonances decaying into quark-antiquark pairs in proton-proton collisions at $\sqrt{s}=13\TeV$}

\date{\today}

\abstract{A search for narrow vector resonances decaying into quark-antiquark
pairs is presented.  The analysis is based on data
collected in proton-proton collisions at \mbox{$\sqrt{s}=13\TeV$} with the CMS detector at the LHC, 
corresponding to an integrated luminosity of 35.9\fbinv.
The hypothetical resonance is produced with sufficiently high transverse
momentum that its decay products are merged into a
single jet with two-prong substructure.
A signal would be identified as a peak over a smoothly falling background in
the distribution of the invariant mass of the jet, using 
novel jet substructure techniques.
No evidence for such a resonance is observed within the mass
range of 50--300\GeV.  Upper limits at 95\% confidence
level are set on the production cross section, and presented in a mass-coupling parameter space.
The limits further constrain simplified
models of dark matter production involving a mediator interacting between
quarks and dark matter particles through a vector or axial-vector
current.  
In the framework of these models, 
the results are the most sensitive to date, extending for the first time the search region to masses
below 100\GeV.
}

\hypersetup{%
pdfauthor={CMS Collaboration},%
pdftitle={Search for low mass vector resonances decaying into quark-antiquark pairs in proton-proton collisions at sqrt(s) = 13 TeV},%
pdfsubject={CMS},%
pdfkeywords={CMS, physics, vector resonances}}

\maketitle

\section{Introduction}

Many extensions of the standard model (SM) predict the existence of new resonances that couple to quarks 
(q)~\cite{Anchordoqui:2008di,Cullen:2000ef,ref_diquark,ref_qstar,Baur:1989kv,ref_axi,Chivukula:2011ng,ref_coloron,Han:2010rf,ref_gauge,ref_rsg}. 
The first searches for such particles 
were reported by the UA1~\cite{Albajar:1988rs} 
and UA2~\cite{Alitti:1990kw,Alitti:1993pn} experiments using 
$\sqrt{s}=630\GeV$ collisions at the CERN $\mathrm{Sp\bar{p}S}$, and 
were extended to larger values of resonance masses by the 
CDF~\cite{Abe:1989gz,Abe:1993it,Abe:1995jz,Abe:1997hm,Aaltonen:2008dn} 
and D0~\cite{Abazov:2003tj} experiments using $\sqrt{s}=1.8$ and $1.96\TeV$ collisions 
at the Fermilab Tevatron.
At the CERN LHC, the searches in proton-proton (pp) collisions at $\sqrt{s} = 7$, 8 and $13\TeV$ performed by the 
ATLAS~\cite{Aad:2010ae,Aad:2011aj,Aad:2011fq,ATLAS:2012pu,Aad:2014aqa,ATLAS:2015nsi,Aaboud:2017yvp}
and 
CMS~\cite{Khachatryan:2010jd,Chatrchyan:2011ns,CMS:2012yf,Chatrchyan:2013qha,Khachatryan:2015sja,Khachatryan:2015dcf,Khachatryan:2016ecr,Sirunyan:2016iap} 
Collaborations have mostly focused on the production of heavy particles.
For resonance masses below 1\TeV, 
the sensitivity is limited by high trigger thresholds 
and by the large expected backgrounds, 
notably from SM events consisting of jets produced through the strong 
interaction, referred to here as QCD multijet events.

These
difficulties can be avoided by an approach focused
on the events where at least one high transverse momentum (\pt) jet
from initial-state radiation (ISR) is produced in association with a
light resonance decaying into a $\PQq\PAQq$ pair. 
The ISR requirement provides enough energy in
the event to satisfy the trigger, either by the ISR jet or by the
resonance itself.  The minimum \pt of the resonance considered
in this search is sufficiently high that the hadronization products of
the daughter quarks merge and are reconstructed as a single, large-radius jet.
The only previous search in this topology to place constraints on resonance masses 
below 300\GeV was by the CMS Collaboration, applying this technique to data collected at the LHC in 2015~\cite{Sirunyan:2017dnz}.

In the current paper, the results of a search for leptophobic vector 
resonances (\PZpr) decaying to quark-antiquark pairs in pp collisions at
$\sqrt{s} = 13\TeV$ are reported, using data collected by the CMS detector in 2016, 
corresponding to an integrated luminosity of $35.9\fbinv$.
The search is performed by looking for a narrow resonance peak in the
continuous jet mass distribution.
The analysis exploits a new substructure variable that is decorrelated from the jet mass
and \pt and preserves the shape of the jet mass distribution used in the search. 
The jet is required to have the two-prong substructure expected from the signal.
The dominant background
from SM QCD multijet production is estimated from a signal-depleted control
region created by inverting the substructure requirement.  The signal
yield is extracted by simultaneously fitting the signal and control
regions, while requiring that the ratio of QCD components in each region is
described by a smooth two-dimensional function of jet mass and \pt. 
The W+jets and Z+jets background components are estimated
from simulation and the top quark background contribution is obtained from simulation 
corrected with scale factors derived from a \ttbar-enriched control sample.

Results are interpreted within the framework of a leptophobic vector
resonance model, and are also used to set limits on the existence of
generic vector-like resonances decaying into
quarks~\cite{Dobrescu:2013coa}.  
Limits are also set in the context of a simplified model 
of dark matter (DM) production at the LHC, 
in which the mediators couple only to quarks and DM particles~\cite{Abercrombie:2015wmb}.
\label{sec:intro}

\section{CMS detector}

The central feature of the CMS apparatus is a superconducting solenoid
of 6\unit{m} internal diameter, providing a magnetic field of
3.8\unit{T}. Within the solenoid volume are a silicon
pixel and strip tracker, a lead tungstate crystal electromagnetic
calorimeter (ECAL), and a brass and scintillator hadron calorimeter
(HCAL), each composed of a barrel and two endcap sections. Forward
calorimeters extend the pseudorapidity ($\eta$)
coverage provided by the barrel and endcap detectors. Muons are
detected in gas-ionization chambers embedded in the steel flux-return
yoke outside the solenoid.  

Events are selected using a two-tiered trigger system~\cite{Khachatryan:2016bia}. 
The first level, composed
of custom hardware processors, uses information from the calorimeters and muon detectors
to select events of interest in a time interval of less than 4\mus. The
second level, known as the high-level trigger (HLT), consists of a farm of processors running a
version of the full event reconstruction software optimized for fast processing, and further reduces the
event rate from around 100 kHz to less than 1 kHz, before data storage.

A more detailed description of the CMS
detector, together with a definition of the coordinate system used and
the relevant kinematic variables, can be found
in Ref.~\cite{Chatrchyan:2008zzk}.
\label{sec:detector}

\section{Event simulation and selection}

\subsection{Simulated samples}

Simulated samples of the \PZpr resonance decaying into
a quark-antiquark pair are generated at leading order (LO) with the
\MGvATNLO 2.2.3 generator~\cite{Alwall:2014hca} with up to 3 extra jets in
matrix element calculations.  The
dominant SM backgrounds arise from multijet and \PW/\PZ+~jets processes.
These backgrounds are simulated at LO using the \MGvATNLO generator with the MLM
matching~\cite{Alwall:2007fs} between jets from matrix element calculations 
and from parton showers, while the \POWHEG 2.0~\cite{POWHEG} generator 
at next-to-leading order (NLO) precision is used to
model the subdominant contribution from pair and single top quark production.  All signal and
background generators are interfaced with \PYTHIA{}
8.212~\cite{Sjostrand:2014zea}, with the CUETP8M1 underlying event tune~\cite{Khachatryan:2015pea}, to simulate 
parton showering and hadronization effects.  The generated
events are further processed through a \GEANTfour~\cite{Agostinelli:2002hh} simulation of the CMS detector.
The parton distribution function (PDF) set {NNPDF3.0}~\cite{Ball:2014uwa} is used to produce all simulated samples, 
with the accuracy (LO or NLO) determined by the generator used.
For events containing \PW~and \PZ~bosons, we apply higher-order QCD
and electroweak (EW) \pt dependent corrections to improve
the modeling of the \pt distribution of \PW~and \PZ events,
following Refs.~\cite{Sirunyan:2017hci,Kallweit:2014xda,Kallweit:2015dum,Kallweit:2015fta,Lindert:2017olm}.
The same NLO QCD corrections that are applied
to the \PW~and \PZ simulation are also applied to the signal
simulation. However, since the coupling of the \PZpr mediator differs from that of
the \PZ boson, the equivalent \PZ NLO EW corrections are not applied
to the signal model.

\subsection{Event reconstruction and selection}

The CMS particle-flow (PF) event
algorithm~\cite{Sirunyan:2017ulk} reconstructs and identifies
individual particles with an optimized combination of information from
the various elements of the CMS detector. 
Each particle candidate is classified as either an electron, a muon, a photon, or a charged or neutral hadron.
The energy of photons is obtained directly from the ECAL measurement, 
corrected for zero-suppression effects.
The energy of electrons is determined from a combination of the electron momentum at the primary interaction vertex
as determined by the tracker, the energy of the corresponding ECAL cluster, and the energy sum of all bremsstrahlung photons spatially
compatible with originating from the electron track. 
The energy of muons is obtained from the curvature of the corresponding track. 
The energy of charged hadrons is determined from a combination of their
momentum measured in the tracker and the matching ECAL and HCAL energy
deposits, corrected for zero-suppression effects and for the response
function of the calorimeters to hadronic showers. 
Finally, the energy of neutral hadrons is obtained from the corresponding corrected ECAL
and HCAL energy. The missing transverse momentum vector is defined as 
the negative vectorial sum of the transverse momenta of all the particles identified in the event, 
and its magnitude is referred to as $\cmsptmiss$.

The PF candidates are clustered into jets using the anti-\kt
algorithm~\cite{Cacciari:2008gp,Cacciari:2011ma}.
Jets are clustered with distance parameters of 0.4 and 0.8, and are referred to as AK4 and AK8 jets, respectively.
To mitigate the impact of particles arising from additional proton-proton interactions 
within the same or adjacent bunch crossings (pileup),
weights calculated with the pileup-per-particle identification 
algorithm~\cite{Bertolini:2014bba} are applied to each PF
candidate prior to jet clustering, based on the likelihood of it coming
from the hard-scattering vertex. 
Further corrections are applied to simulated jet energies 
as a function of jet $\eta$ and \pt to match 
the observed detector response~\cite{jec,Khachatryan:2016kdb}.

This search focuses on events in which a high-\pt jet from a merged
$\PZpr \to \PQq\PAQq$ recoils against another high-\pt
ISR jet.  A combination of several online signatures is required for
the trigger selection, all requiring the total hadronic transverse
energy in the event (\HT) or the jet \pt to exceed a certain threshold.  
In addition, soft radiation remnants are removed with the jet trimming technique~\cite{Krohn:2009th} 
before the mass selection, allowing the \HT and jet \pt trigger thresholds to be reduced, and improving the signal acceptance.
To be fully efficient with respect to the trigger requirement, we require at
least one AK8 jet with $\pt > 500\GeV$ and $|\eta| < 2.5$. 
Additional quality criteria are applied to the jets in order to remove
spurious jet-like features originating from isolated noise patterns in
the calorimeters or the tracker.  The efficiency of these jet quality
requirements for signal events is above $99\%$.  
In order to reduce backgrounds from SM EW processes, events are removed
if they contain identified and isolated electrons, muons, or taus with $\pt > 10\GeV$
and $|\eta| < 2.5$, $2.4$, or $2.3$, respectively, according to the isolation criteria in~\cite{Sirunyan:2017hci}.

\begin{figure}[ht]
\centering
\includegraphics[width=0.75\linewidth] {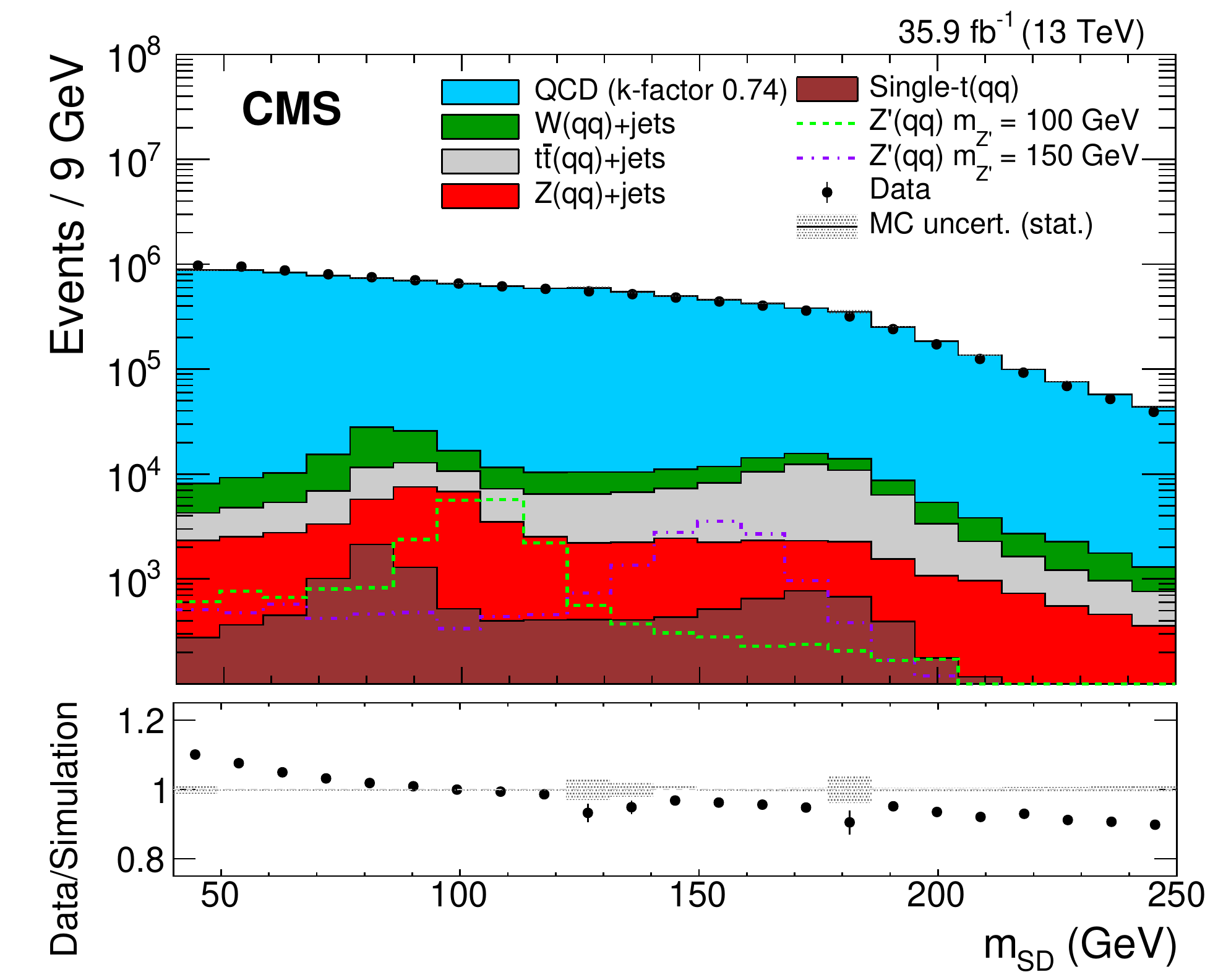}
\caption{Distributions of data (points) and simulated backgrounds (histograms), of the leading $\pt$ jet
soft-drop mass after the jet
kinematic selection.  Dashed lines
illustrate the signal contribution for different $\PZpr$ boson
masses.  The multijet processes (QCD) dominate the background
component, with subdominant contributions from inclusive SM $\PW$, $\PZ$,
and $\ttbar$ and single top quark processes.
The QCD simulation is corrected by an overall factor of 0.74 to match the data yield.
\label{fig:control1}
}
\end{figure}

In the subsequent offline analysis, the most energetic jet in the event is assumed to 
correspond to the $\PZpr \to \PQq\PAQq$ system, and is reconstructed as a single AK8 jet. 
The search is performed using the distribution of the jet mass groomed
with the soft-drop algorithm ($m_\text{SD}$), which is an extension of the modified 
mass drop tagger~\cite{Dasgupta:2013ihk,Larkoski:2014wba} that 
removes soft and wide-angle radiation produced by
parton shower activity, pileup interactions, and the underlying event from the jet.
Jets are groomed using the parameters $z_\text{cut} = 0.1$ and $\beta = 0$.
Here, $z_\text{cut}$ specifies subleading the energy fraction relative to the whole jet at which jet declustering into subjet pairs is stopped.  
The parameter $\beta$ adds additional angular requirements on the jet declustering.  For 
$\beta = 0$, these requirements are neglected, and approximately the same fraction of energy is groomed away
regardless of the initial jet energy~\cite{Larkoski:2014wba}.
The soft-drop grooming reduces the jet mass for QCD background jets 
when large masses arise from soft gluon radiation.
In contrast, the jet mass for merged 
$\PZpr \to \PQq\PAQq$ and $\PW/\PZ \to \PQq\PAQq$~jets
comes from the kinematic distributions of the
decay, and is largely unchanged by grooming.
Figure~\ref{fig:control1} shows the  
distributions of $m_\text{SD}$
for data and simulation, after the jet kinematic selection.

In this paper, the dimensionless scaling variable $\rho$~\cite{Dasgupta:2013ihk,Dolen:2016kst}, 
defined as $\rho=\ln(m_\text{SD}^2/\pt^2)$,
is used in the characterization of
the correlation of jet substructure variables with the jet mass and
\pt.  For QCD jets, the distribution of $\rho$
is approximately invariant under a change of jet \pt,
in the region where perturbative contributions dominate and scale as ($m_\text{SD}/\pt$).
This property does not hold in two regimes: in the
low mass region below $\rho\approx -6$, 
where non-perturbative effects are large and scale as ($1/m_\text{SD}$) instead,
and in the high mass region above $\rho\approx -2$.
The departure from $\rho$ invariance in the latter case 
arises because the cone size of the AK8 jets 
is insufficient to provide complete containment at high masses.
Consequently, only events in the range $-5.5 < \rho < -2.0$ are considered. 
This requirement is fully efficient for
the \PZpr~boson signal and roughly translates to a $m_\text{SD}$
range from 25\GeV to 185\GeV at $\pt = 500\GeV$.

In addition to the jet mass, the observable $N_2^{1}$~\cite{Moult:2016cvt} 
is used to discriminate the two-prong structure of the
jets from the $\PZpr \to \PQq\PAQq$ decay  
from the hadronization products of single light quarks or gluons, 
which are overwhelmingly one-prong.
This jet substructure variable 
is defined from a combination of generalized 
energy correlation functions ${_{v}e_n}$,
sensitive to correlations of $v$ pairwise angles among $n$-jet constituents~\cite{Moult:2016cvt}.
In particular, the 2-point ($_{1}e_{2}$) and 3-point ($_{2}e_{3}$) correlation functions are defined as:
\begin{align}
_{1}e_{2} &= \sum_{1\leq i < j \leq n}{z_{i}z_{j}\Delta R_{ij}}~, \\
_{2}e_{3} &= \sum_{1\leq i < j < k\leq n}{z_{i}z_{j}z_{k} \min \{\Delta R_{ij}\Delta R_{ik}, \Delta R_{ij}\Delta R_{jk}, \Delta R_{ik}\Delta R_{jk} \}}~,
\end{align}
where $z_i$ represents
the energy fraction of the constituent $i$ in the jet and $\Delta R_{ij}$
is the angular separation between constituents $i$ and $j$.
For a two-prong structure, signal jets have a stronger 2-point correlation
than a 3-point correlation. 
The discriminant variable $N_2^{1}$ is then constructed via the ratio:
\begin{equation}
\quad N_{2}^{1}= \frac{_{2}e_{3}}{(_{1}e_{2})^{2}}~.
\end{equation}

The energy correlation functions are
computed from the jet constituents after the soft-drop grooming has
been applied, thereby reducing their dependence on the jet mass and
$\pt$~\cite{Moult:2016cvt}.

The $N_2^{1}$ observable has excellent performance in discriminating
two-prong signal jets from multijet QCD background
jets~\cite{Moult:2016cvt}.  However, $N_2^{1}$ and similar
variables are correlated with the jet mass and \pt.  A selection based
on $N_2^{1}$ would distort the jet mass distribution, 
with the amount of distortion depending on the
\pt of the jet. 
This would make the search for a resonant peak in the jet mass distribution, 
over a large range of \pt, particularly challenging.

\begin{figure}[ht]
\centering
\includegraphics[width=0.75\textwidth] {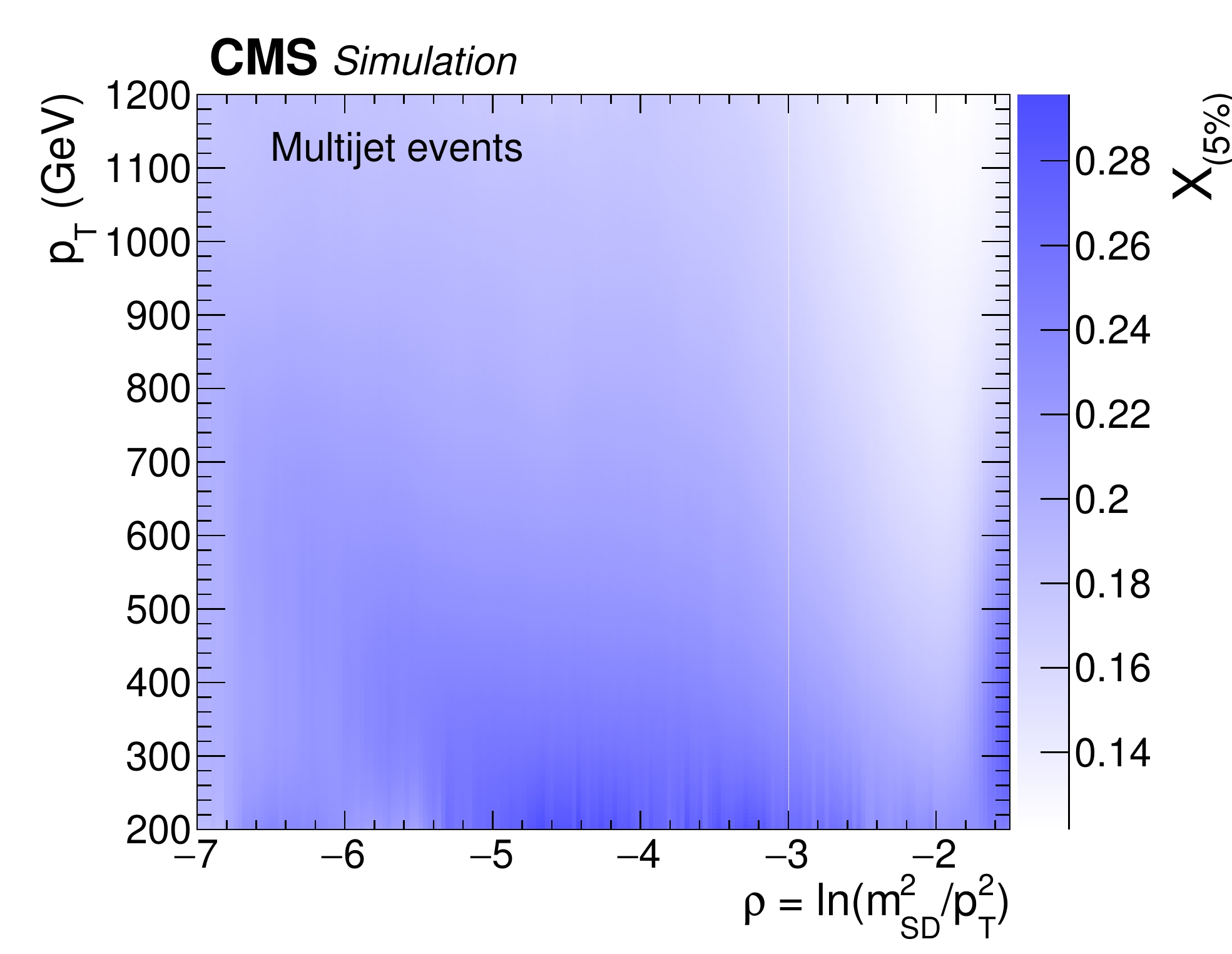}
\caption{The distribution of $X_{(5\%)}$ used to define the $N_2^{1,\text{DDT}}$ variable, corresponding
to the 5\% quantile of the $N_2^{1}$ distribution in simulated multijet events.
The distribution is shown as a function of the jet
$\rho$ and $\pt$ and smoothed using a kNN approach~\cite{Dudani:1976}. 
The $N_2^{1}$ distribution is mostly insensitive to the jet $\rho$ and $\pt$ in the
kinematic phase space considered for this analysis ($-5.5 < \rho <                                                                                                                
-2.0$). Residual correlations in simulation are corrected by applying 
a decorrelation procedure that yields the $N_2^{1,\text{DDT}}$
variable.
\label{fig:N2ddt}
}
\end{figure}

The key feature of our approach is that the application of the substructure 
requirement preserves the shape of the soft-drop jet mass distribution.
Improving on the decorrelation procedure proposed in Ref.~\cite{Dolen:2016kst}, 
we apply a DDT (designed decorrelated tagger) transformation of $N_2^{1}$ to  $N_2^{1,\text{DDT}}$. 
It is defined as $N_2^{1,\text{DDT}}(\rho,\pt)$~$\equiv$~$N_2^{1}(\rho,\pt)-X_{(5\%)}(\rho,\pt)$,
where $X_{(5\%)}$ is derived from the simulated $N_2^{1}$ distribution and illustrated in Fig.~\ref{fig:N2ddt}.
We require events to pass the $N_2^{1,\text{DDT}}(\rho,\pt) < 0$ selection, such that
we select a fixed 5\% of QCD multijet events independent of $\rho$ and \pt.
The distribution of $X_{(5\%)}$ is smoothed using a distance weighted k-nearest neighbor (kNN) approach~\cite{Dudani:1976}.
The chosen percentile maximizes the sensitivity to the \PZpr~boson signal.

\begin{figure}[ht]
\centering
\includegraphics[width=0.75\linewidth] {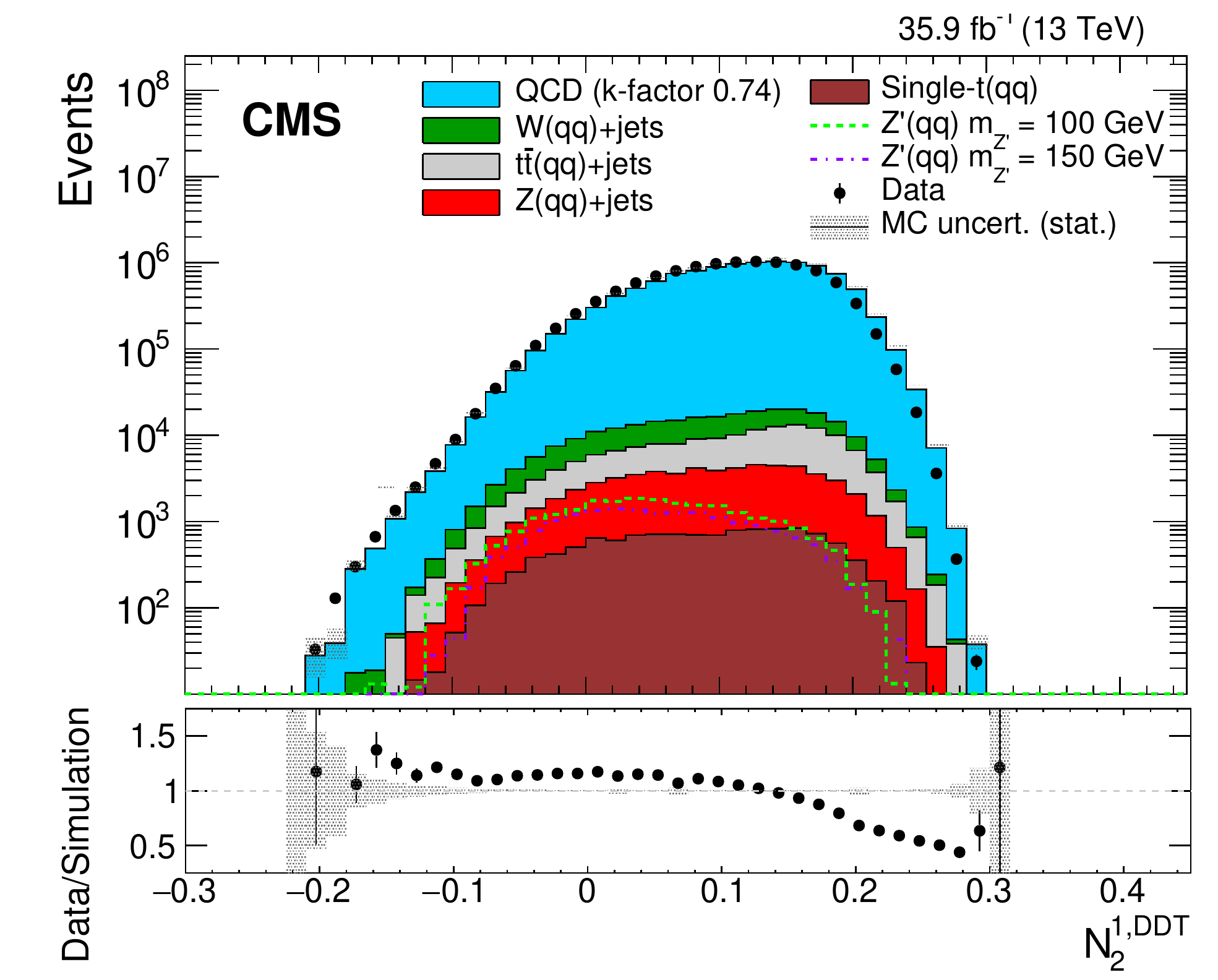}
\caption{ Distributions of data (points) and simulated backgrounds (histograms), 
of the $N_2^{1,\text{DDT}}$ variable for the leading $\pt$ jet after the
kinematic selection.  Dashed lines
illustrate the signal contribution for different $\PZpr$ boson 
masses.  The multijet processes (QCD) dominate the background
component, with subdominant contributions from inclusive SM $\PW$, $\PZ$,
and $\ttbar$ and single top quark processes. 
The QCD simulation is corrected by an overall factor of 0.74 to match the data yield.
\label{fig:control2}
}
\end{figure}

The distributions of $N_2^{1,\text{DDT}}$ for
data and simulation are shown in Fig.~\ref{fig:control2} after the
jet $\pt > 500\GeV$ requirement. 
Since there is a visible disagreement between simulation and data,
the multijet background is estimated from data, as described in the next section.
Additional distributions of kinematic observables for data and simulation
are available in Appendix~\ref{app:suppMat}.

\label{sec:evsel}

\section{Background estimate}

The search is performed by looking for a resonance in the soft-drop
mass distribution over background contributions dominated by QCD multijet events
and smaller contributions from $\PW(\PQq'\PAQq)$+jets,
$\PZ(\PQq\PAQq)$+jets, and top quark background processes.

To model the background contribution from pair and single top quark production
we utilize simulation with data-driven corrections 
based on a dedicated control region. 
This region has the same
kinematic requirements as the signal region but with the muon veto inverted. 
The muon is selected using dedicated muon triggers and is required to
have $\pt > 100\GeV$ and $|\eta| < 2.1$ and to be in the opposite
hemisphere to the selected AK8 jet.  To enrich the \ttbar
contribution and reduce the multijet contamination, at least one AK4
jet with $\pt > 50\GeV$ is required to pass the b-tagging medium
selection based on the combined secondary vertices version-2 algorithm~\cite{Chatrchyan:2012jua},
which identifies AK4 jets that originate from the
hadronization of b quarks.  Separate scale factors correct 
the overall top quark background normalization and
the $N_2^\text{1,DDT}$ efficiency for mistagging jets
from top quark decays. These scale factors are
$\mathrm{SF}^{\text{\ttbar}}_{\mathrm{norm}} = 0.75 \pm 0.10$ 
and $\mathrm{{SF}^{t}_{\mathrm{mistag}}} = 0.83 \pm 0.03$, respectively.

Subdominant backgrounds arising from resonant SM
processes ($\PW$/$\PZ$+~jets) are estimated from simulations that include
corrections to the shape and normalization from higher order NLO QCD
and EW calculations. Additional data-to-simulation corrections
for the jet mass shapes and $N_2^\text{1,DDT}$ tagging efficiencies
are applied to the simulation. These corrections are evaluated from a \ttbar control region rich in merged hadronic
$\PW$~bosons, as further explained below.

\begin{figure}[ht]
\centering
\includegraphics[width=0.45\textwidth] {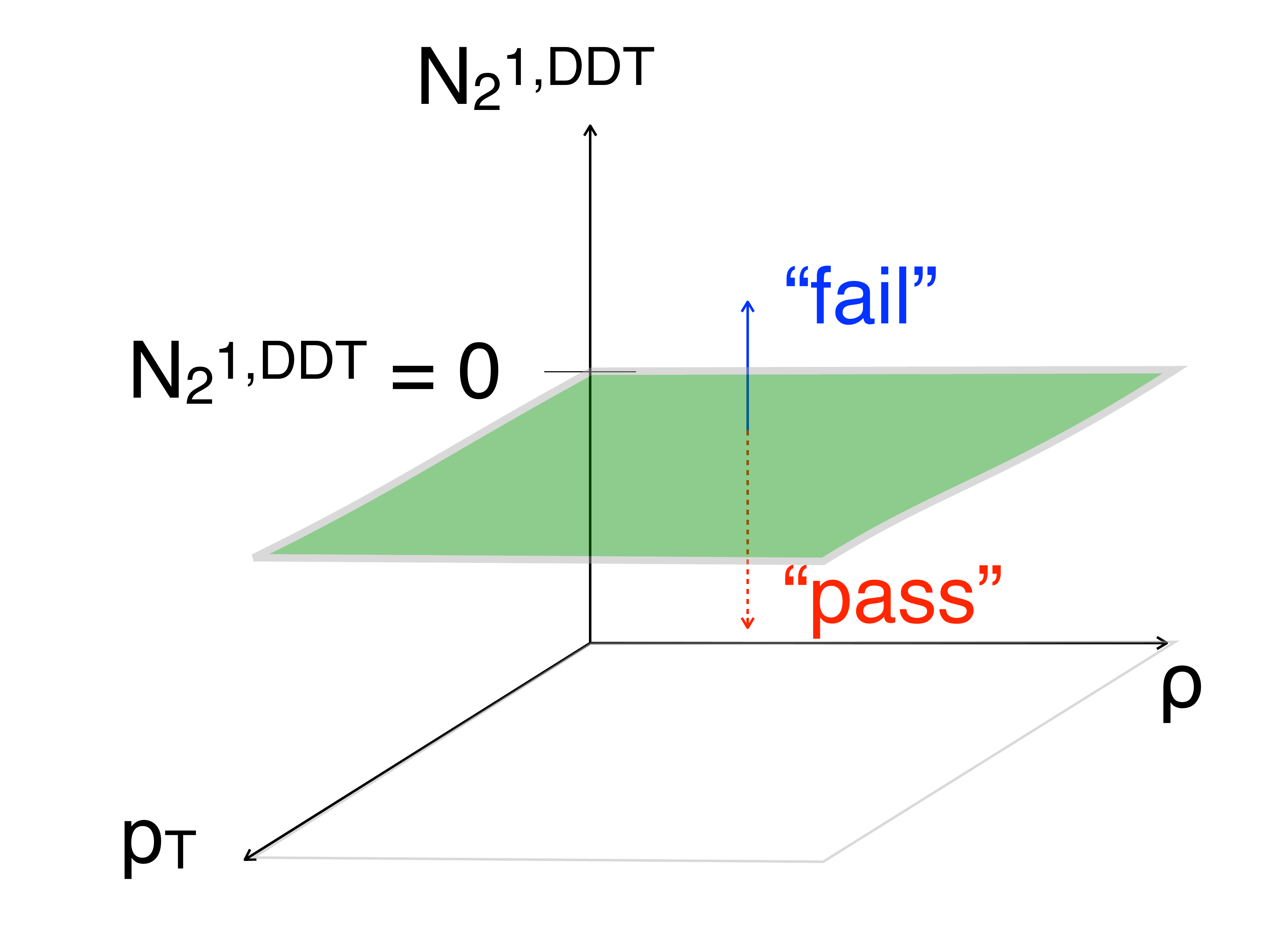} \hfil
\includegraphics[width=0.45\textwidth] {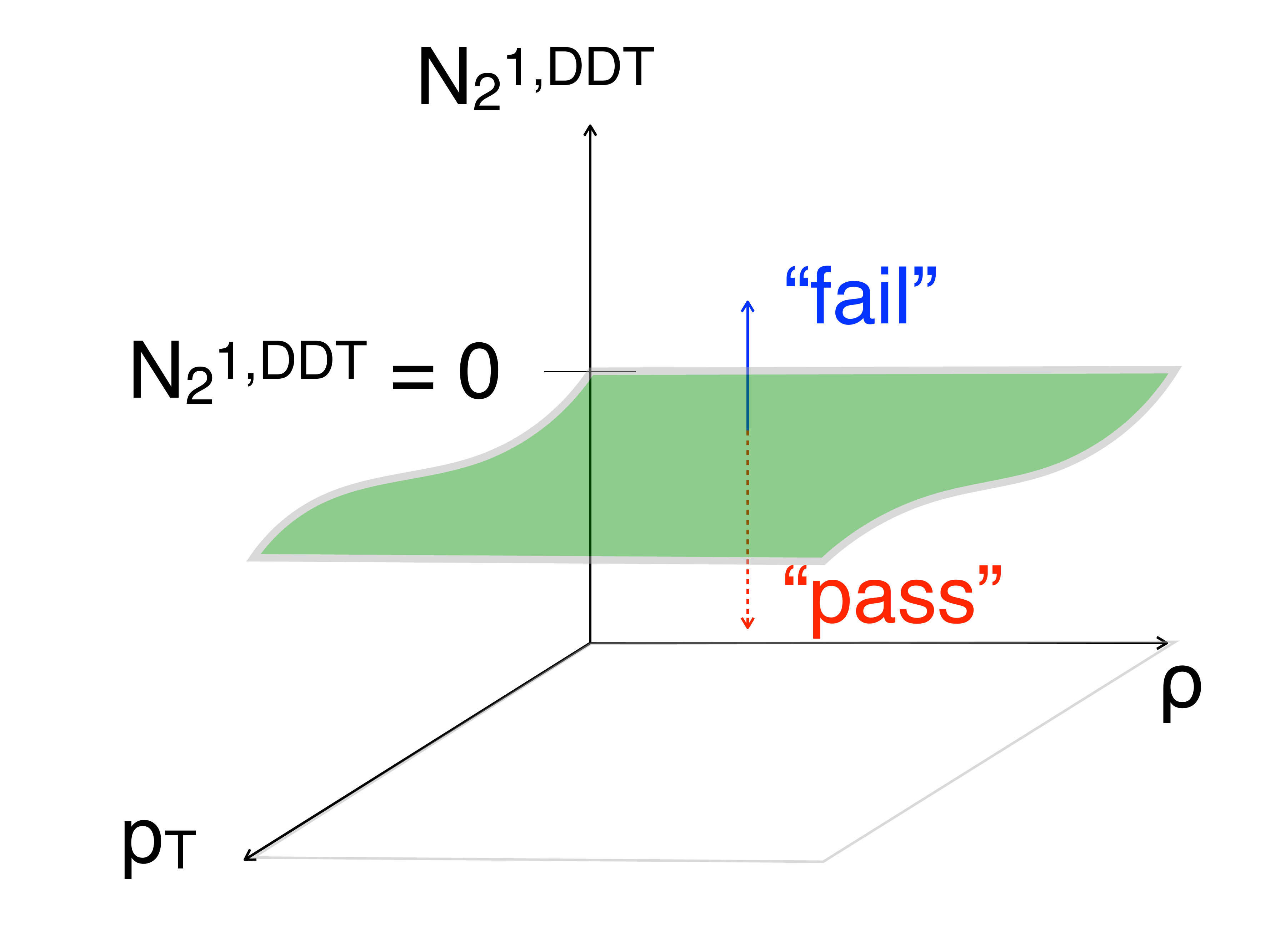}
\caption{
A schematic of the background estimation method.
The pass-to-fail ratio, $R_{\rm p/f}(\rho(m_\text{SD},\pt))$, is defined from the events passing and failing the $N_2^\text{1,DDT}$ selection.
The variable $N_2^\text{1,DDT}$ is constructed so that, for simulated multijet events, $R_{\rm p/f}$ is constant (left).
To account for residual differences between data and simulation, 
$R_{\rm p/f}$ is extracted by performing a two-dimensional fit to data in ($ \rho, \pt$) space (right).
\label{fig:rhalphstrat2}
}
\end{figure}

We estimate the main QCD multijet event background 
by taking advantage of the decorrelation of $N_2^\text{1,DDT}$ from $\rho$ and \pt.  
The fraction of events passing the $N_2^\text{1,DDT}$ selection is, by construction, a constant 5\% in simulated multijet events.
The decorrelation ensures that the 
events passing and failing the selection
have the same shape of the QCD jet mass distribution, 
and their ratio, the ``pass-to-fail ratio'' $R_{\rm p/f}$, 
is constant for simulated multijet events.
The prediction of events passing the selection can then be expressed as:
\begin{equation}
\label{eq:qcd_prediction}
n^{\mathrm{QCD}}_\mathrm{pass} (m_\text{SD}, \pt) = R_{\rm p/f}(\rho(m_\text{SD}, \pt),
\pt) \, n^{\mathrm{QCD}}_\mathrm{fail}(m_\text{SD}, \pt)~,
\end{equation}
where $n^{\mathrm{QCD}}_\mathrm{pass}$ and $n^{\mathrm{QCD}}_\mathrm{fail}$ are
the number of passing and failing events in a given $m_\text{SD}$, \pt bin.
This procedure is illustrated schematically in
Fig.~\ref{fig:rhalphstrat2}. 
Since the distribution of $\rho$ is
expected to be invariant under a change of \pt, 
$R_{\rm p/f}$ is parametrized
as a function of $\rho$, which is in turn
expressed as a function of $m_\text{SD}$ and \pt.

Owing to residual differences between data and simulation, 
the correction $R_{\rm p/f}(\rho,\pt)$ is allowed to deviate from a constant. 
The deviation is modeled by expanding $R_{\rm p/f}(\rho,\pt)$ into a polynomial series in orders of $\rho$ and $\pt$:
\begin{align}
\label{eq:tf}
R_{\rm p/f}(\rho, \pt) &=  \epsilon_{\mathrm{QCD}}(1 + a_{01} \pt + a_{02} \pt^2 + \cdots \\
&+ (a_{10}+ a_{11} \pt + a_{12} \pt^2 + \cdots)\rho \nonumber \\
& + (a_{20}+ a_{21} \pt + a_{22} \pt^2+\cdots)\rho^2  +\cdots ) \nonumber.
\end{align}

The coefficients $\epsilon_{\mathrm{QCD}}$ and $a_{k\ell}$
have no external constraints but are
determined from a simultaneous fit to the data events passing and
failing the substructure requirement, together with the signal yield.
The number of required coefficients in the fit is determined with a Fisher $F$-test on
data~\cite{ref:ftest} by iteratively adding polynomial orders. The optimum choice is
found to be of fourth order in $\rho$ and third order in \pt.
The fact that $R_{\rm p/f}$ varies slowly across the $m_\text{SD}$--\pt
domain is essential, since it allows one to estimate the background under a
narrow signal resonance based on the events across the whole jet mass range.
\label{sec:bkg}

\section{Systematic uncertainties}

Uncertainties in the multijet background arise from the fit parameter uncertainties in the pass-to-fail ratio fit 
described in Eq.~(\ref{eq:tf}). 
The uncertainties in the top quark background normalization (10\%) and $N_{2}^\text{1,DDT}$ mistag
(2\%) scale factors are
propagated to the signal extraction through the fit.

\begin{figure}[hbtp]
\begin{center}
\includegraphics[width=0.45\textwidth]{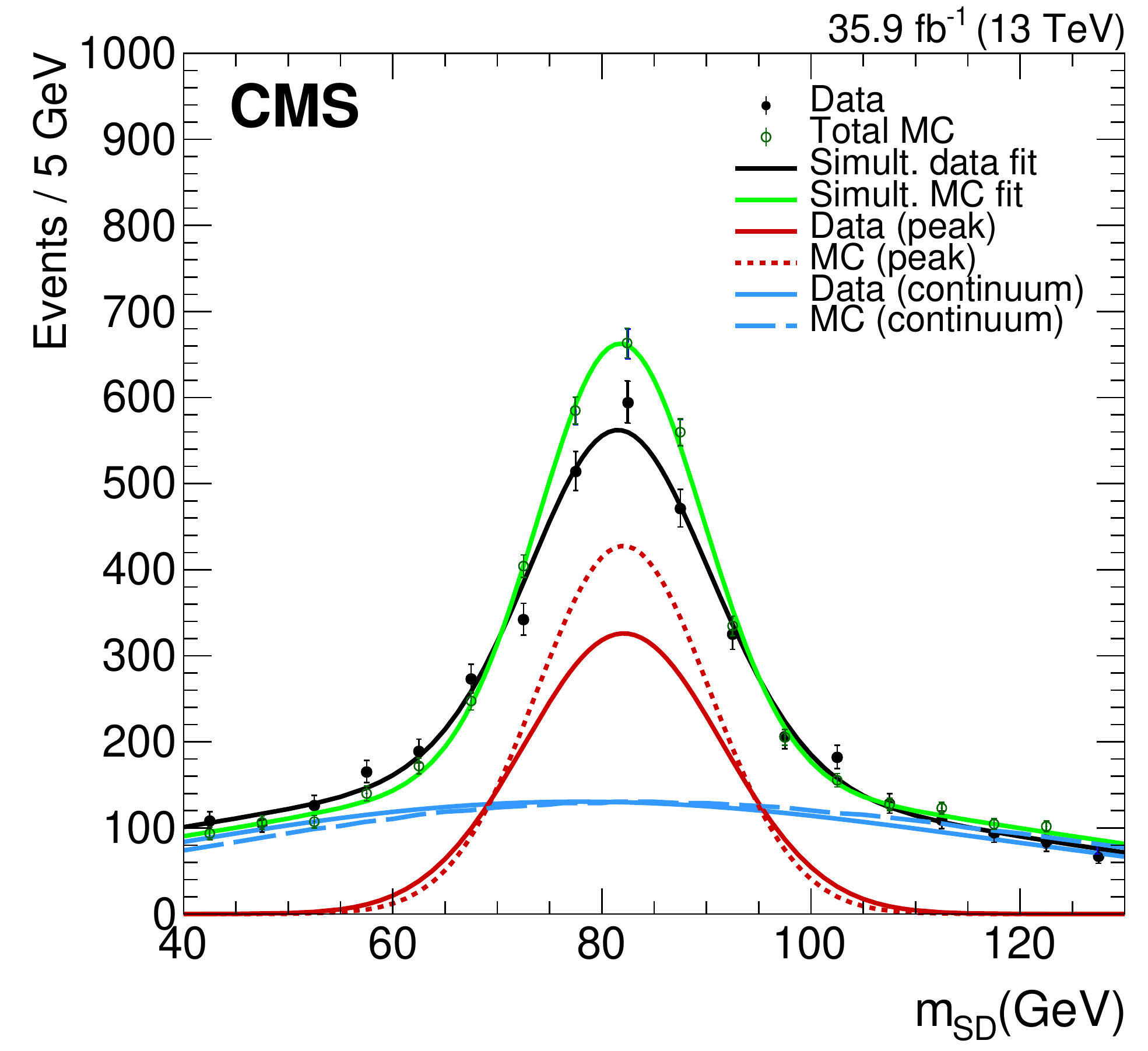} \hfil
\includegraphics[width=0.45\textwidth]{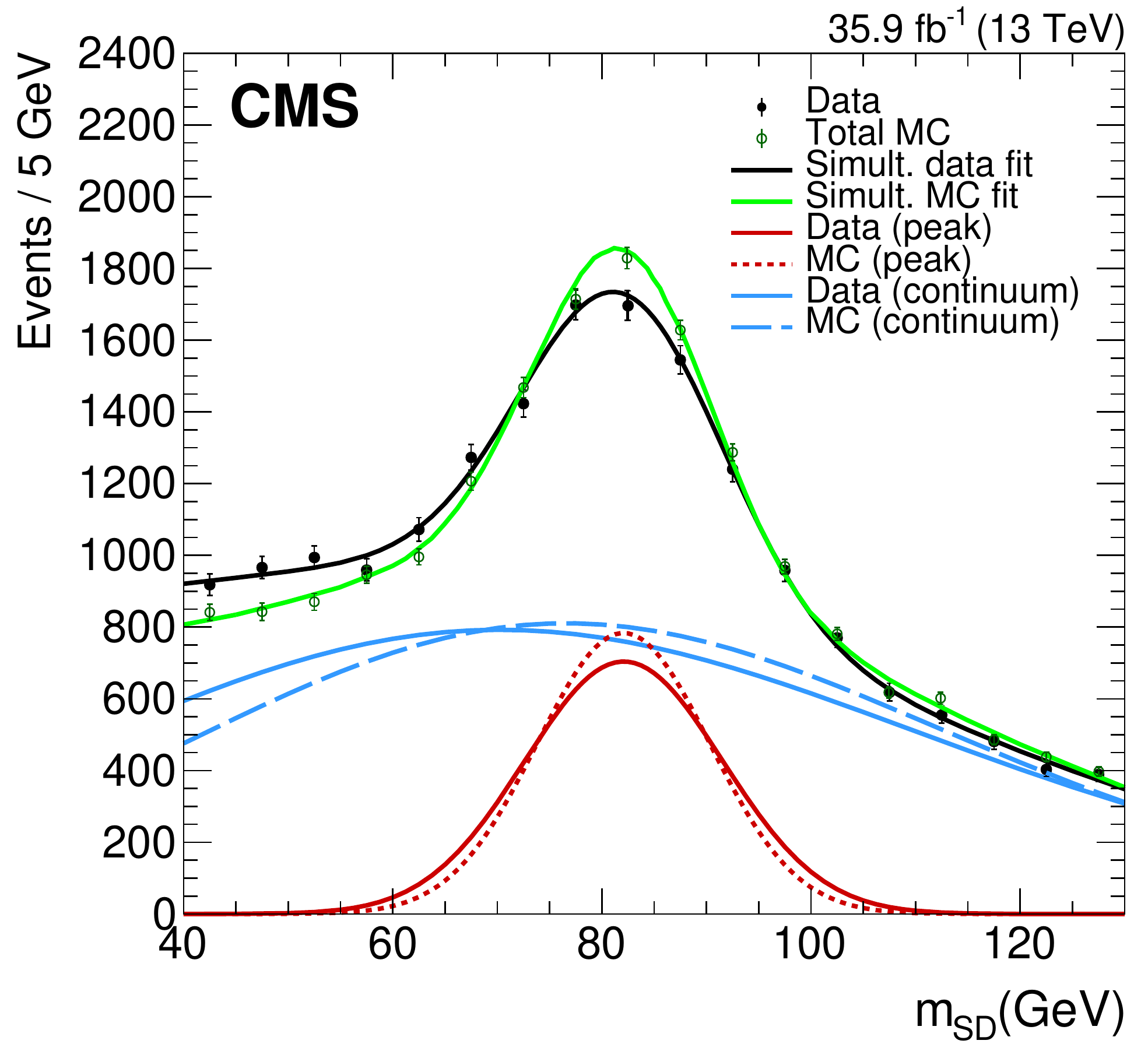}
\caption{ Soft-drop jet mass distributions that pass (left) and fail (right) the $N_{2}^\text{1,DDT}$ selection in the semileptonic $\ttbar$ sample. Results of fits to
data and simulation are shown. }
\label{fig:ttbar}
\end{center}
\end{figure}

The systematic effects for the shapes and normalization of
the \PW, \PZ backgrounds, and signal components are strongly correlated since they are 
affected by similar systematic mismeasurements.  We constrain the jet mass
scale, the jet mass resolution, and the $N_2^\text{1,DDT}$ selection
efficiency using a sample of merged \PW~boson jets in semileptonic
\ttbar events in data.  In this region, events are required to have an energetic 
muon with $\pt > 100\GeV$, $\cmsptmiss > 80\GeV$,
a high-\pt AK8 jet with $\pt > 200\GeV$, and a b-tagged AK4 jet separated from the AK8 jet by
$\Delta R > 0.8$.  Using the same
$N_2^\text{1,DDT}$ requirement described above, we define samples
with events that pass and fail the selection for merged \PW~boson jets in data and
simulation, shown in Fig.~\ref{fig:ttbar}.
A simultaneous fit to the two samples is performed in order to extract
the selection efficiency of a merged \PW~jet in simulation and in data.
We measure the data-to-simulation scale
factor for the $N_2^\text{1,DDT}$ selection to be $ 0.88 \pm 0.10 $.
The mass scale between data and simulation is found to be $ 1.10 \pm 0.05 $. 
The jet mass resolution data-to-simulation scale factor is
measured to be $ 1.14 \pm 0.06 $.  
These scale factors determine the initial distributions of the jet mass
for the \PW, \PZ boson, and signal and they are further constrained in the fit to data 
because of the presence of the \PW~and~\PZ resonances in the jet mass
distribution.  
To account for potential deviations due to missing 
higher-order corrections to the simulated boson \pt distributions, 
uncertainties are assumed in the \PW~and~\PZ boson yields that are \pt-dependent. 
An additional systematic uncertainty 
is included to account for potential differences between the \PW~and~\PZ boson 
higher-order corrections.
Finally, uncertainties associated to the jet energy resolution~\cite{jec}, 
trigger efficiency, lepton veto efficiency, 
and the integrated luminosity determination~\cite{lumi} 
are also applied to
the \PW, \PZ boson, and \PZpr boson signal yields.
A quantitative summary of the systematic effects considered is listed in Table~\ref{tab:sys}.

\begin{table}[hbtp]
\begin{center}
\topcaption{Summary of the systematic uncertainties for signal and background processes and their relative size.
The symbol $^\triangle$ denotes uncertainties decorrelated per $\pt$ bin in the 500--1000\GeV range.
The symbol $^\dagger$ denotes a shape uncertainty in the peaking SM $\PW$~and $\PZ$ boson backgrounds and $\PZpr$ boson signal shape. A long dash (\NA) indicates that the uncertainty does not apply.}
\begin{tabular}{ lcccc } 
\hline 
Systematic source & Multijet & \PZpr & \PW/\PZ & \ttbar \\ 
\hline 
Lepton veto efficiency & \NA & 0.5\% & 0.5\% & \NA\\
Jet mass scale$^\dagger$ & \NA & 0.5\% & 0.5\% & \NA\\
Jet mass scale (\pt dependence)~$^\dagger$$^\triangle$ & \NA & 0.5--2\%\y & 0.5--2\%\y & \NA\\
Trigger efficiency & \NA & 2\% & 2\% & \NA\\
Top quark mistag rate & \NA & \NA & \NA & 2\%\\
Integrated luminosity & \NA & 2.5\% & 2.5\% & \NA\\
Multijet fit parameters & \x1--3\% & \NA & \NA & \NA\\ 
$N_{2}^\text{1,DDT}$ selection efficiency & \NA & 9\% & 9\% & \NA\\
Top quark background normalization & \NA & \NA & \NA & 10\%\\
Jet energy resolution$^\dagger$ & \NA & 10\% & 10\% & \NA\\
NLO QCD corrections & \NA & 10\% & 10\% & \NA\\
NLO EW corrections$^\triangle$ & \NA & \NA & \x15--35\% & \NA \\
NLO EW \PW/\PZ decorrelation  & \NA & \NA & \x\x5--15\% & \NA\\
\hline
\end{tabular}
\label{tab:sys}
\end{center}
\end{table}

To validate the robustness of the fit, we perform a goodness-of-fit test
and bias tests using pseudo-experiments and injecting a simulated signal, for different values of
\PZpr boson mass. No significant bias is observed.
As a further test of fit robustness, we split the region failing the $N_2^{1,DDT}$ selection 
into two smaller regions mimicking the passing and failing regions in the signal extraction fit. 
The mimicked passing-like region corresponds to a background efficiency of 
60--65\% and the mimicked failing-like region corresponds to an efficiency of 65--100\%.
We repeat our background estimation procedure on this selection as if the 60--65\% efficiency region  
were the passing region.
We find negligible biases in the fitted signal strength.
\label{sec:sys}

\section{Results}

\begin{figure}[hbtp]
\begin{center}
\includegraphics[width=0.45\textwidth]{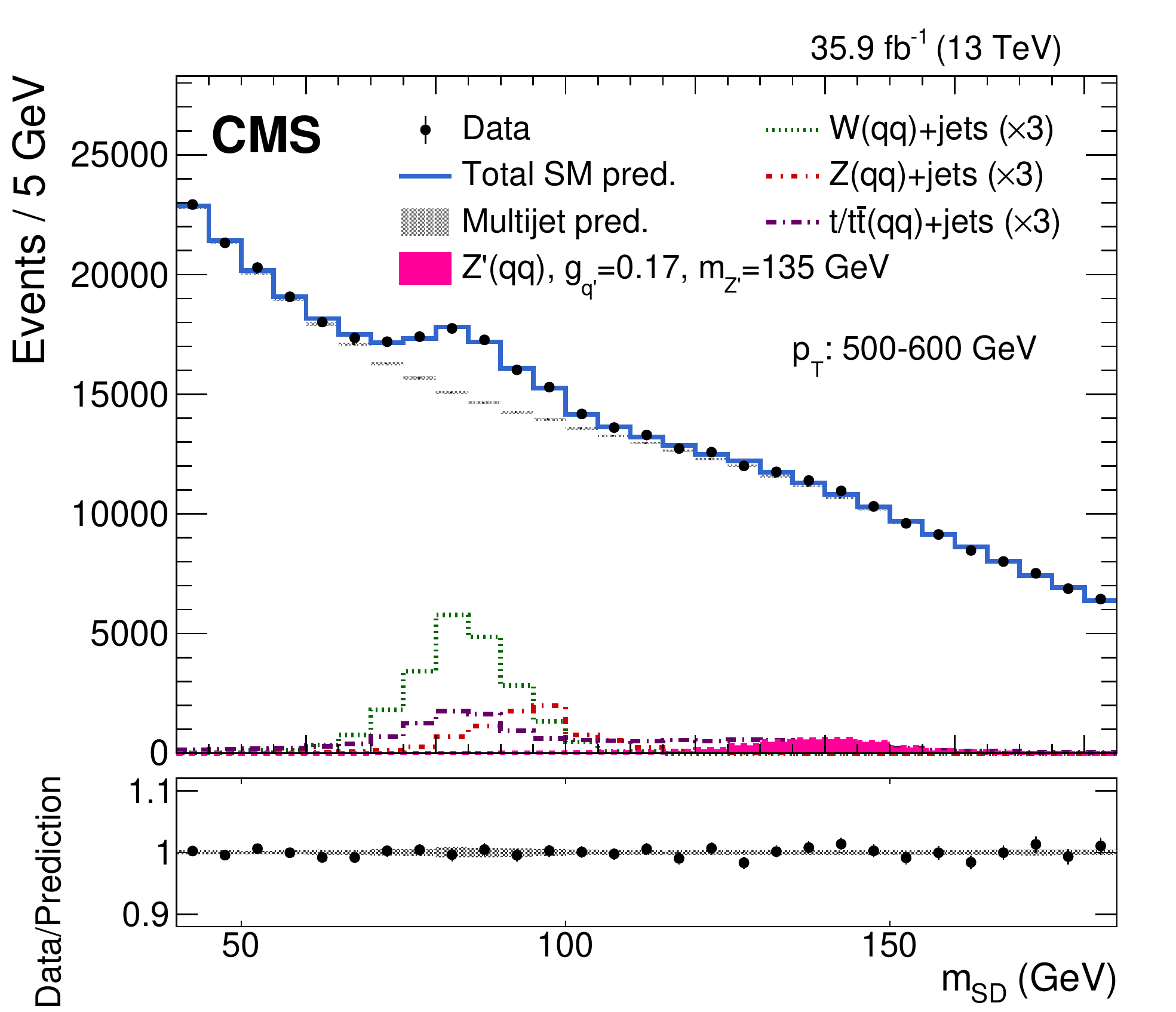} \hfil
\includegraphics[width=0.45\textwidth]{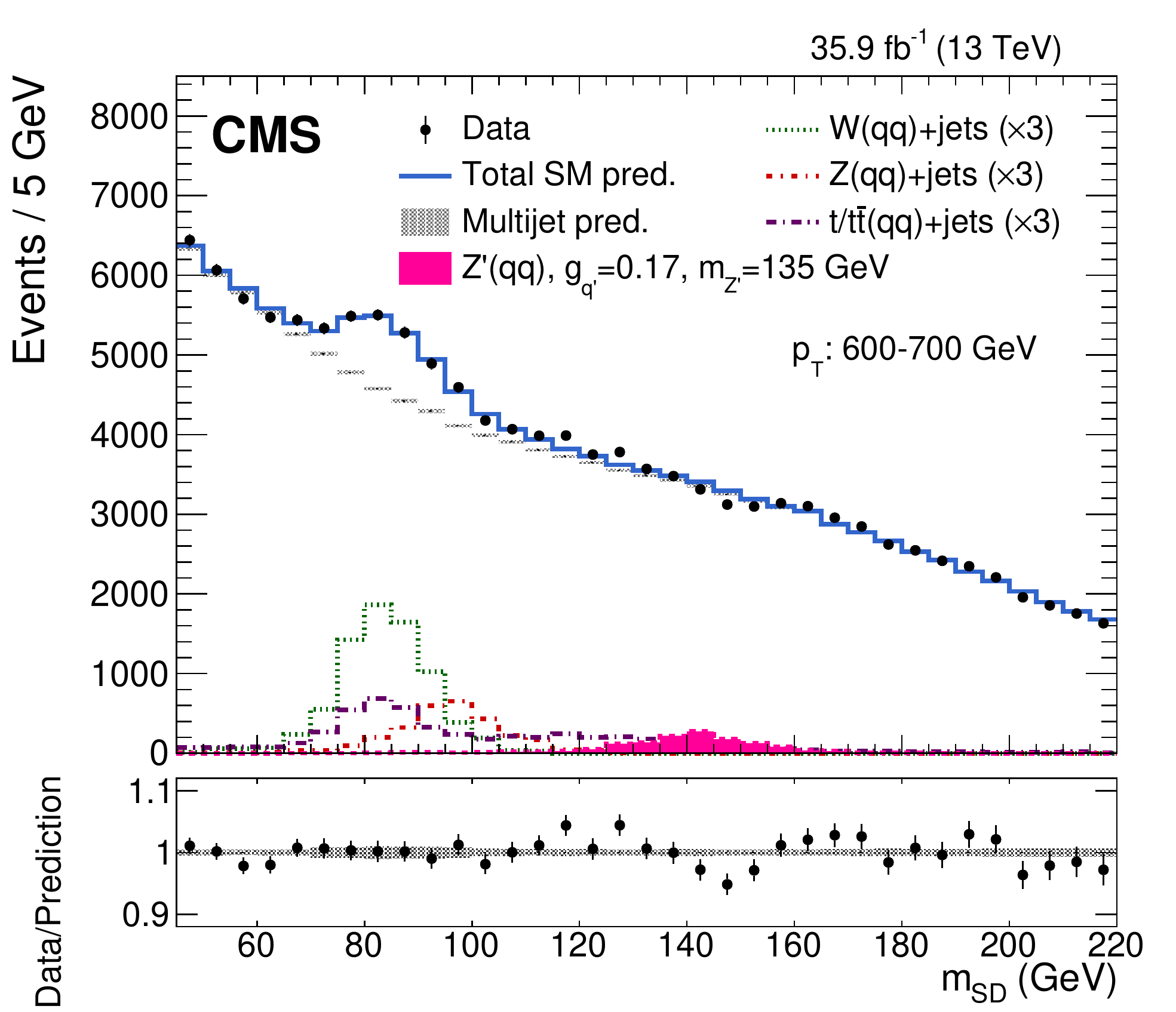}\\  
\includegraphics[width=0.45\textwidth]{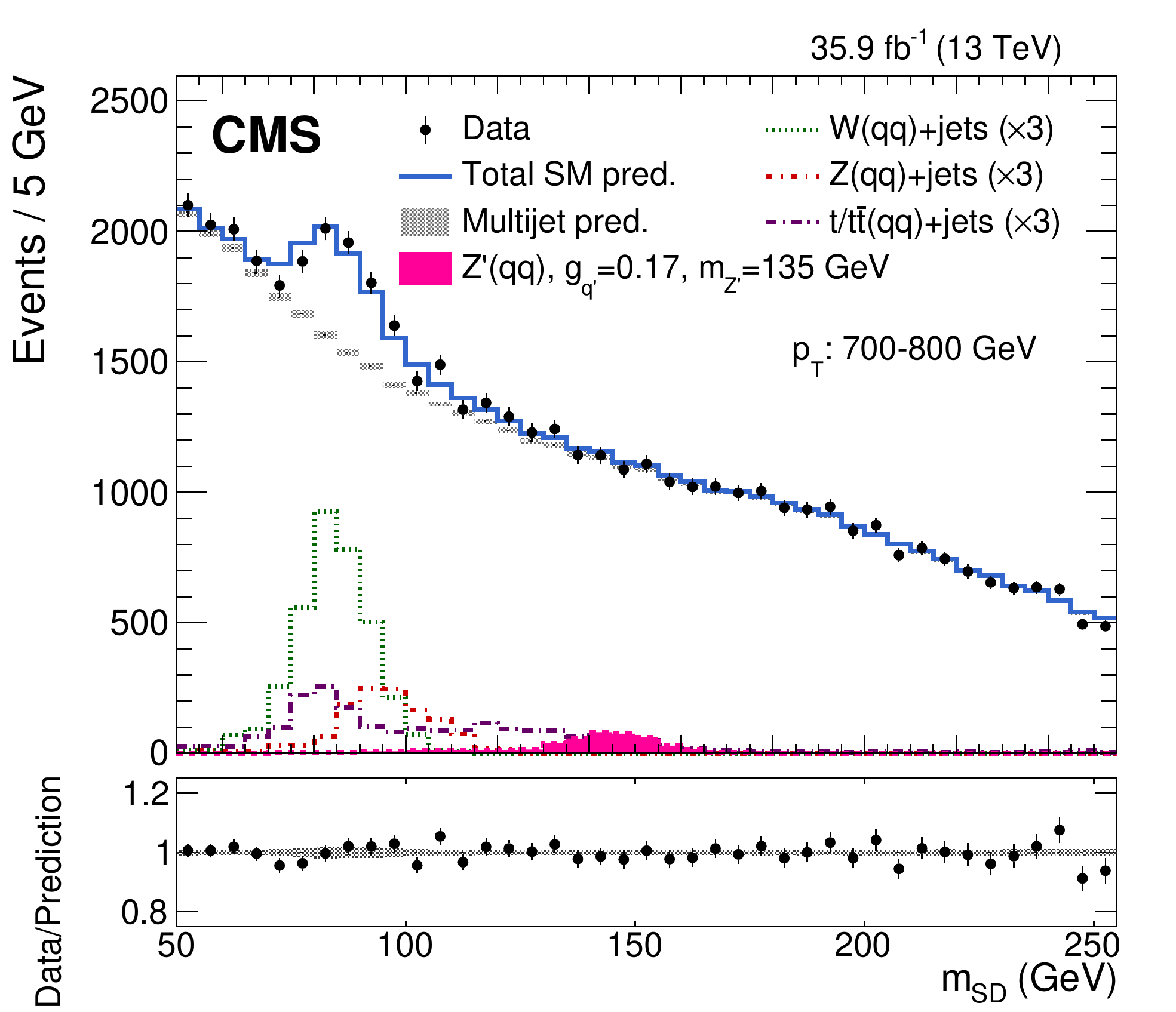} \hfil
\includegraphics[width=0.45\textwidth]{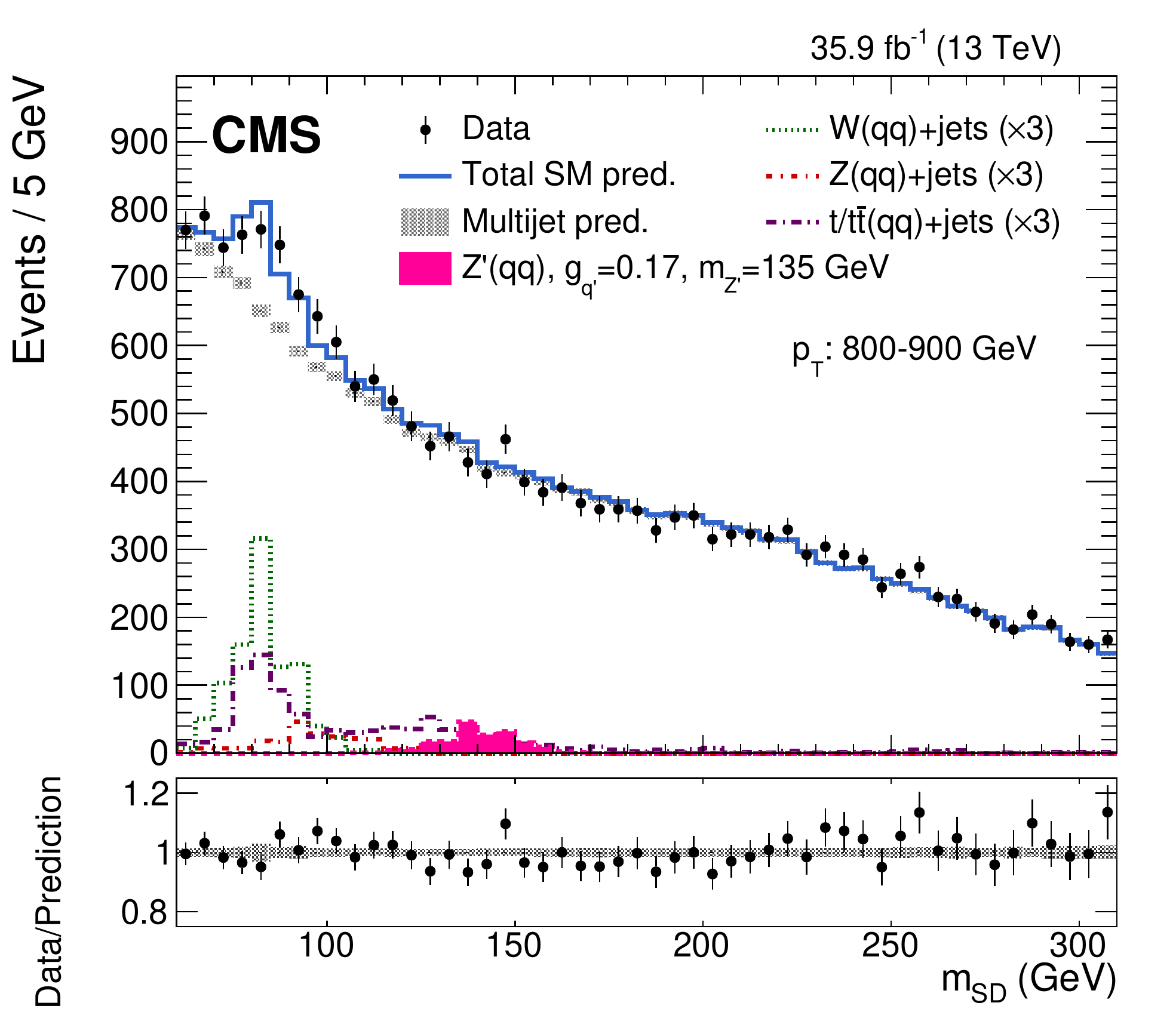}\\
\includegraphics[width=0.45\textwidth]{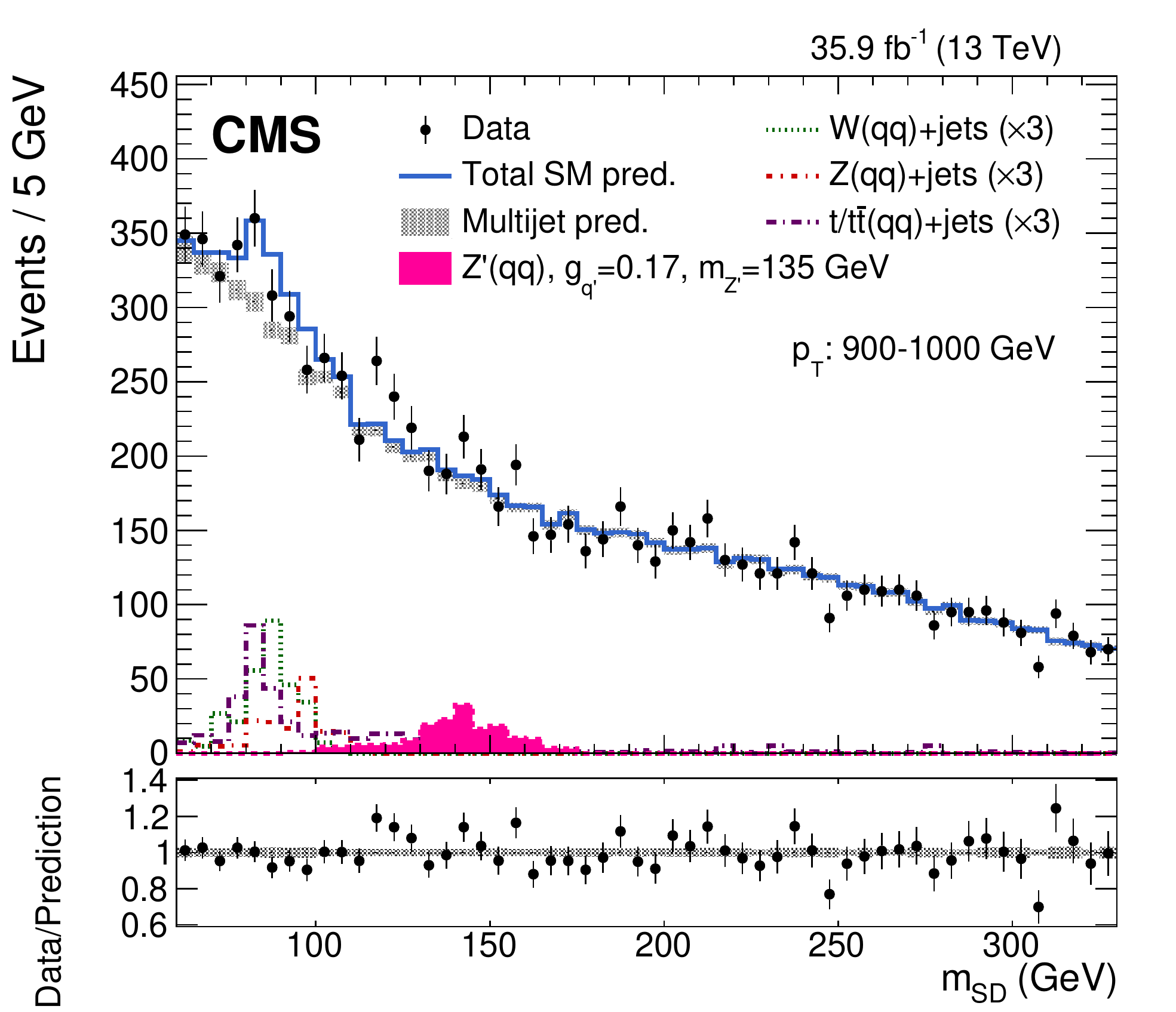}
\end{center}
\begin{center}
\caption{Soft-drop jet mass distribution for the different \pt ranges of the fit from 500 to 1000\GeV. 
Data are shown as black points. 
The multijet background prediction, including uncertainties, is shown by the shaded bands. 
Contributions from the \PW~and~\PZ~boson, and top quark background processes 
are shown, scaled up by a factor of 3 for clarity.
A hypothetical \PZpr boson signal at a mass of 135\GeV is also indicated. 
In the bottom panel, the ratio of the data to the background prediction, including uncertainties, is shown. 
The scale on the x-axis differs for each \pt range due to the kinematic selection on $\rho$.}
\label{fig:money}
\end{center}
\end{figure}

We combine the estimates of the various SM background processes and 
search for a potential signal from a \PZpr~resonance in the mass range from 50 to 300\GeV.
A binned maximum likelihood fit to the observed shape of the
soft-drop jet mass distribution is performed simultaneously in the passing and failing regions
of five \pt ranges whose boundaries are: 500,~600,~700,~800,~900 and 1000\GeV.
The number of observed events is consistent
with the predicted background from SM processes.
Figure~\ref{fig:money} shows the
soft-drop jet mass distribution for data and measured background contributions in the different \pt ranges 
for a \PZpr mass of 135\GeV;
the \PW~and \PZ~boson contributions are clearly visible in the data.
The $m_{\text{SD}}$ distribution for data in the combined \pt ranges 
is available in Appendix~\ref{app:suppMat}.

\begin{figure}[hbtp]
\begin{center}
\includegraphics[width=0.45\textwidth]{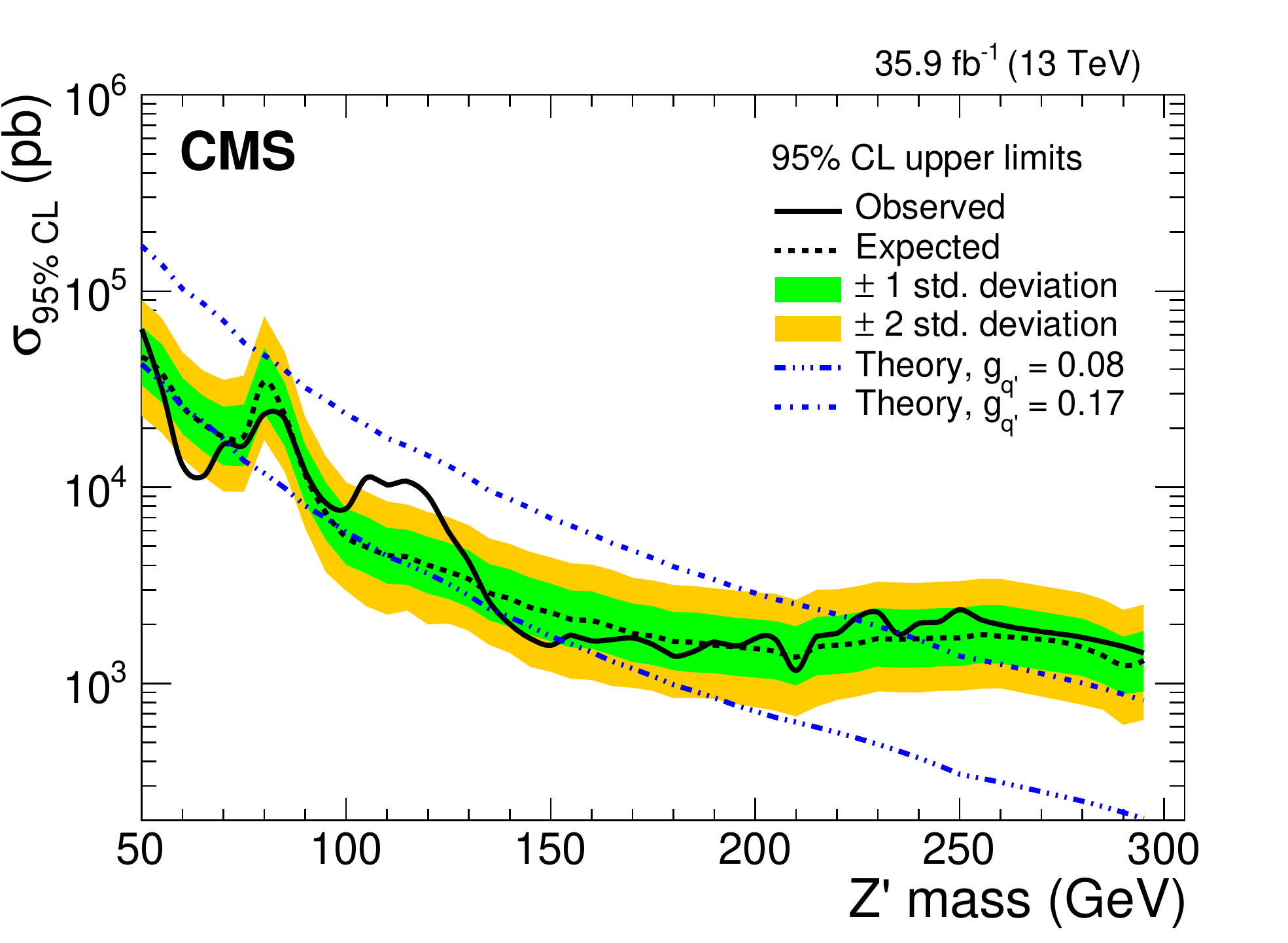} \hfil
\includegraphics[width=0.45\textwidth]{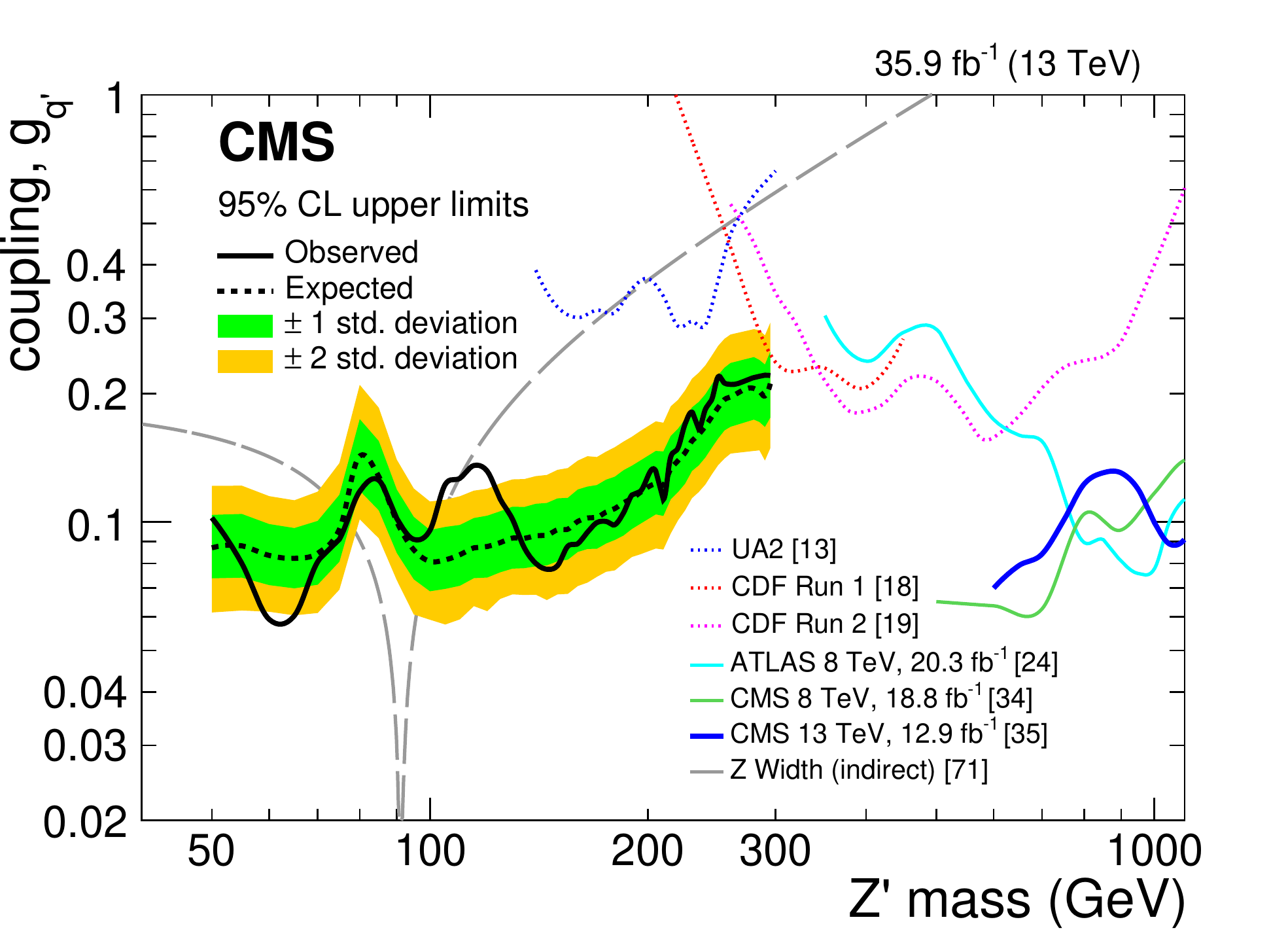}
\caption{
The 95\% CL upper limits on the $\PZpr$ boson production cross section compared to theoretical cross sections (left) and
on the quark coupling $g_{\rm q^{\prime}}$ as a function of resonance mass for a leptophobic $\PZpr$ resonance that only couples to quarks~(right).
The observed limits (solid), expected limits (dashed) and their variation at the 1 and 2 standard deviation levels (shaded bands) are shown.
Limits from other relevant searches and an indirect constraint on a potential $\PZpr$ signal from the SM $\PZ$~boson width~\cite{Dobrescu:2014fca} are also shown. 
\label{fig:limits}    
}
\end{center}
\end{figure}

The results are
interpreted in terms of 95\% confidence level (CL) upper limits on the production cross
section. Upper limits are computed using the modified frequentist
approach for confidence levels (CL$_s$); taking the profile likelihood as
the test statistic~\cite{CLS2,CLS1} in the asymptotic approximation~\cite{Cowan:2010js}.
They are shown as a function of the resonance mass in Fig.~\ref{fig:limits}~(left), where
they are compared to cross sections for 
a model of a leptophobic \PZpr resonance with 
quark coupling $g_{\rm q^\prime}$ value of either~$0.17$ or $0.08$ that are close to our current sensitivity.
Systematic uncertainties are treated as nuisance
parameters, which are modeled with log-normal priors 
and profiled over in the limit calculations.
The maximum local observed $\text{p}$-value corresponds to $2.9$ standard deviations from the
background-only expectation at a \PZpr boson mass of 115\GeV, and the global significance,
calculated over the probed mass range~\cite{Gross:2010qma},
corresponds to approximately $2.2$ standard deviations.

Upper limits on the signal cross section are translated into the coupling $g_{\rm q^{\prime}}$ as a function of \PZpr boson mass, 
related to the \PZpr coupling convention of Ref.~\cite{Dobrescu:2013coa} by 
$g_{\rm q^{\prime}} = g_{\rm B}/6$.
Coupling values above the solid curves are excluded. In Fig.~\ref{fig:limits}~(right), we show previous results
from UA2, CDF, ATLAS and CMS experiments.  Indirect constraints from the
hadronic \PZ boson partial width measurement and limits from the UA2 and CDF experiments are interpreted
from~\cite{Dobrescu:2013coa}.  

\begin{figure}[hbtp]
\begin{center}
\includegraphics[width=0.75\textwidth]{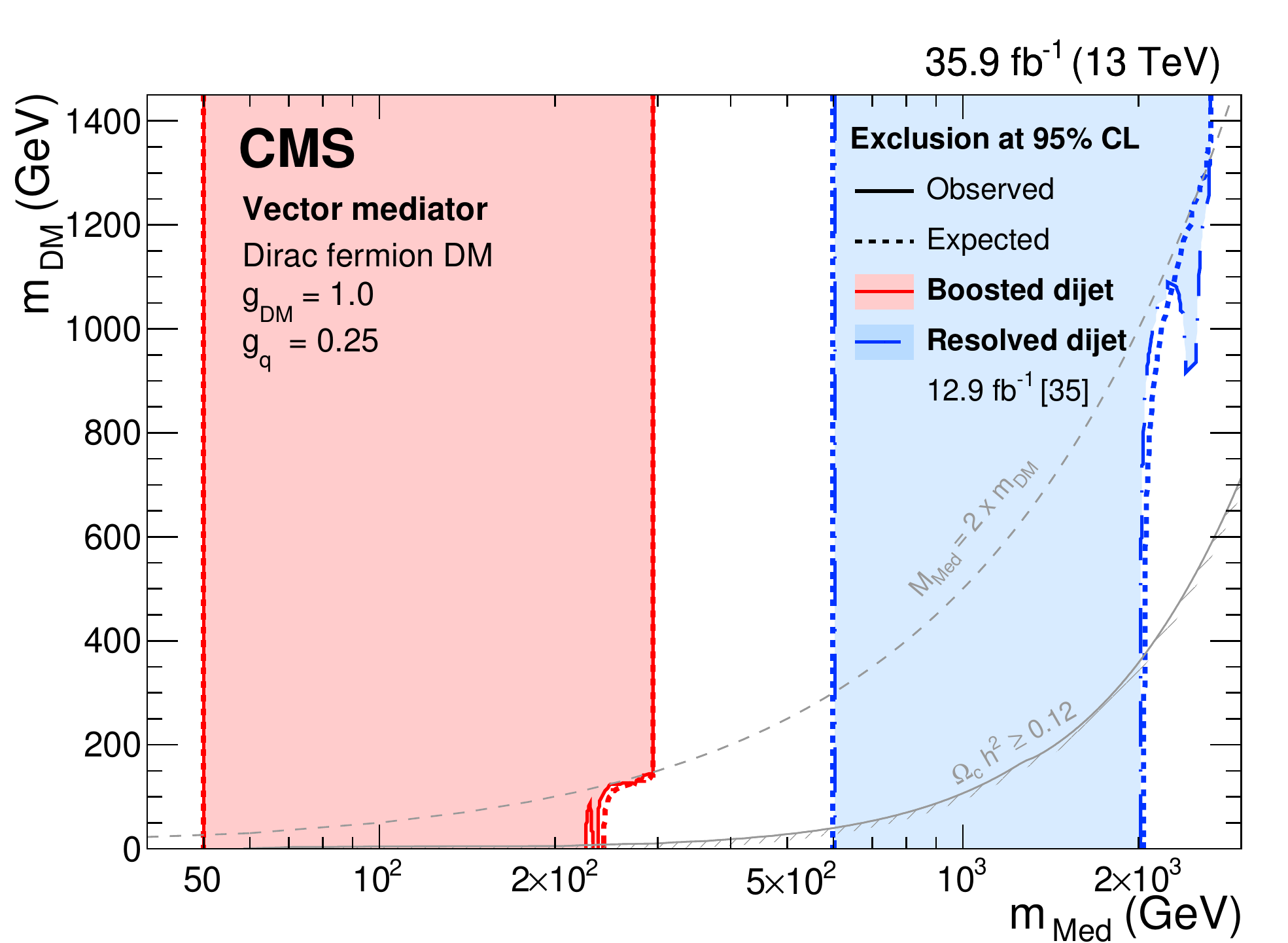}
\caption{
The 95\% CL observed (solid red) and expected (dashed red) 
excluded regions in the plane of dark matter particle mass ($m_{\rm DM}$) vs. mediator mass ($m_{\rm Med}$), for vector mediators. 
A branching fraction of $100\%$ is assumed for a leptophobic vector mediator decaying to dijets.
The exclusion is computed for a quark coupling choice $g_{q} = 0.25$ and for a dark matter coupling $g_{\rm DM} = 1$. 
The excluded regions from the dijet resolved analysis (blue dot dashed lines) using early 2016 data~\cite{Sirunyan:2016iap} are also shown.
Results are compared to constraints from the cosmological relic density of DM (light gray) 
determined from astrophysical measurements~\cite{Spergel:2006hy,Ade:2013zuv} and \textsc{MadDM} version
2.0.6~\cite{Backovic:2013dpa,Backovic:2015cra} as described in Ref.~\cite{Pree:2016hwc}.
}   
\label{fig:DMlimits}    

\end{center}
\end{figure}

The results of this analysis can be used to constrain simplified models of DM. 
Figure~\ref{fig:DMlimits} shows the excluded values at 95\% CL of mediator mass ($m_{\rm Med}$) as a function of the dark matter particle mass ($m_{\rm DM}$) 
for vector mediators, in simplified models that assume a leptophobic mediator that couples only to quarks and DM particles~\cite{Boveia:2016mrp,Abercrombie:2015wmb}. 
Limits are shown for a choice of universal quark coupling $g_{\rm q} = 0.25$ and a DM coupling $g_{\rm DM} = 1.0$.
The difference in limits between axial-vector and vector mediator couplings is small and thus only constraints for the latter coupling scenario are shown.
The excluded range of mediator mass (red) is between 50 and 300\GeV.
The upper bound decreases to 240\GeV when $m_{\rm Med} > 2m_{\rm DM}$, because the branching fraction (BR) to $\PQq\PAQq$ decreases 
as the BR to DM becomes kinematically favorable.
If $m_{\rm Med} < 2m_{\rm DM}$, the mediator cannot decay to DM particles and 
the dijet cross section from the mediator model becomes identical to that in the leptophobic \PZpr model, 
meaning that the limits on the mediator mass in Fig.~\ref{fig:DMlimits} are identical to the limits on the \PZpr mass 
with a coupling $g_{\rm q^{\prime}} = g_{\rm q} = 0.25$. 
For axial-vector mediators, the excluded values of mediator mass are expected to be identical 
to the excluded values in Fig.~\ref{fig:DMlimits} when $m_{\rm DM} > m_{\rm Med}/2$ or $m_{DM}=0$, 
with differences only expected in the transition region $m_{\rm Med} \simeq 2m_{\rm DM}$.
Additional limits (blue) in Fig.~\ref{fig:DMlimits} come from traditional dijet searches~\cite{Sirunyan:2016iap}.

\section{Summary}

A search for a vector resonance ($\PZpr$) decaying into a quark-antiquark pair and reconstructed as a single jet 
has been presented, using a data set comprising proton-proton collisions at \mbox{$\sqrt{s}=13\TeV$}, corresponding to an integrated
luminosity of 35.9\fbinv.
Novel substructure techniques are employed to identify a jet containing a $\PZpr$ boson candidate
over a smoothly falling soft-drop jet mass distribution in data.
No significant excess above the SM prediction is observed, 
and 95\% confidence level upper limits are set on the $\PZpr$ boson coupling to quarks, $g_{\rm q^{\prime}}$, as a function of the
$\PZpr$ boson mass. 
Coupling values of $g_{\rm q^{\prime}} > 0.25$ are excluded over the $\PZpr$ mass range from
50 to 300\GeV, with strong constraints for masses less than 200\GeV.
The results obtained for masses
from 50 to 100\GeV represent the first 
direct limits to be published in this range.
Limits are set on a simplified model of dark matter mediators 
that only couple to quarks and dark matter particles, 
excluding vector mediators with masses between 50 and 300\GeV, 
and using a universal quark coupling $g_{\rm q} = 0.25$ and a dark matter coupling $g_{\rm DM} = 1.0$.
\label{sec:res}

\begin{acknowledgments}

We congratulate our colleagues in the CERN accelerator departments for the excellent performance of the LHC and thank the technical and administrative staffs at CERN and at other CMS institutes for their contributions to the success of the CMS effort. In addition, we gratefully acknowledge the computing centres and personnel of the Worldwide LHC Computing Grid for delivering so effectively the computing infrastructure essential to our analyses. Finally, we acknowledge the enduring support for the construction and operation of the LHC and the CMS detector provided by the following funding agencies: BMWFW and FWF (Austria); FNRS and FWO (Belgium); CNPq, CAPES, FAPERJ, and FAPESP (Brazil); MES (Bulgaria); CERN; CAS, MoST, and NSFC (China); COLCIENCIAS (Colombia); MSES and CSF (Croatia); RPF (Cyprus); SENESCYT (Ecuador); MoER, ERC IUT, and ERDF (Estonia); Academy of Finland, MEC, and HIP (Finland); CEA and CNRS/IN2P3 (France); BMBF, DFG, and HGF (Germany); GSRT (Greece); OTKA and NIH (Hungary); DAE and DST (India); IPM (Iran); SFI (Ireland); INFN (Italy); MSIP and NRF (Republic of Korea); LAS (Lithuania); MOE and UM (Malaysia); BUAP, CINVESTAV, CONACYT, LNS, SEP, and UASLP-FAI (Mexico); MBIE (New Zealand); PAEC (Pakistan); MSHE and NSC (Poland); FCT (Portugal); JINR (Dubna); MON, RosAtom, RAS, RFBR and RAEP (Russia); MESTD (Serbia); SEIDI, CPAN, PCTI and FEDER (Spain); Swiss Funding Agencies (Switzerland); MST (Taipei); ThEPCenter, IPST, STAR, and NSTDA (Thailand); TUBITAK and TAEK (Turkey); NASU and SFFR (Ukraine); STFC (United Kingdom); DOE and NSF (USA).

\hyphenation{Rachada-pisek} Individuals have received support from the Marie-Curie programme and the European Research Council and Horizon 2020 Grant, contract No. 675440 (European Union); the Leventis Foundation; the A. P. Sloan Foundation; the Alexander von Humboldt Foundation; the Belgian Federal Science Policy Office; the Fonds pour la Formation \`a la Recherche dans l'Industrie et dans l'Agriculture (FRIA-Belgium); the Agentschap voor Innovatie door Wetenschap en Technologie (IWT-Belgium); the Ministry of Education, Youth and Sports (MEYS) of the Czech Republic; the Council of Science and Industrial Research, India; the HOMING PLUS programme of the Foundation for Polish Science, cofinanced from European Union, Regional Development Fund, the Mobility Plus programme of the Ministry of Science and Higher Education, the National Science Center (Poland), contracts Harmonia 2014/14/M/ST2/00428, Opus 2014/13/B/ST2/02543, 2014/15/B/ST2/03998, and 2015/19/B/ST2/02861, Sonata-bis 2012/07/E/ST2/01406; the National Priorities Research Program by Qatar National Research Fund; the Programa Severo Ochoa del Principado de Asturias; the Thalis and Aristeia programmes cofinanced by EU-ESF and the Greek NSRF; the Rachadapisek Sompot Fund for Postdoctoral Fellowship, Chulalongkorn University and the Chulalongkorn Academic into Its 2nd Century Project Advancement Project (Thailand); the Welch Foundation, contract C-1845; and the Weston Havens Foundation (USA).

\end{acknowledgments}
\bibliography{auto_generated}

\clearpage
\appendix

\section{Supplementary materials}
\label{app:suppMat}

\begin{figure}[!htb]
\begin{center}
\includegraphics[width=0.45\textwidth]{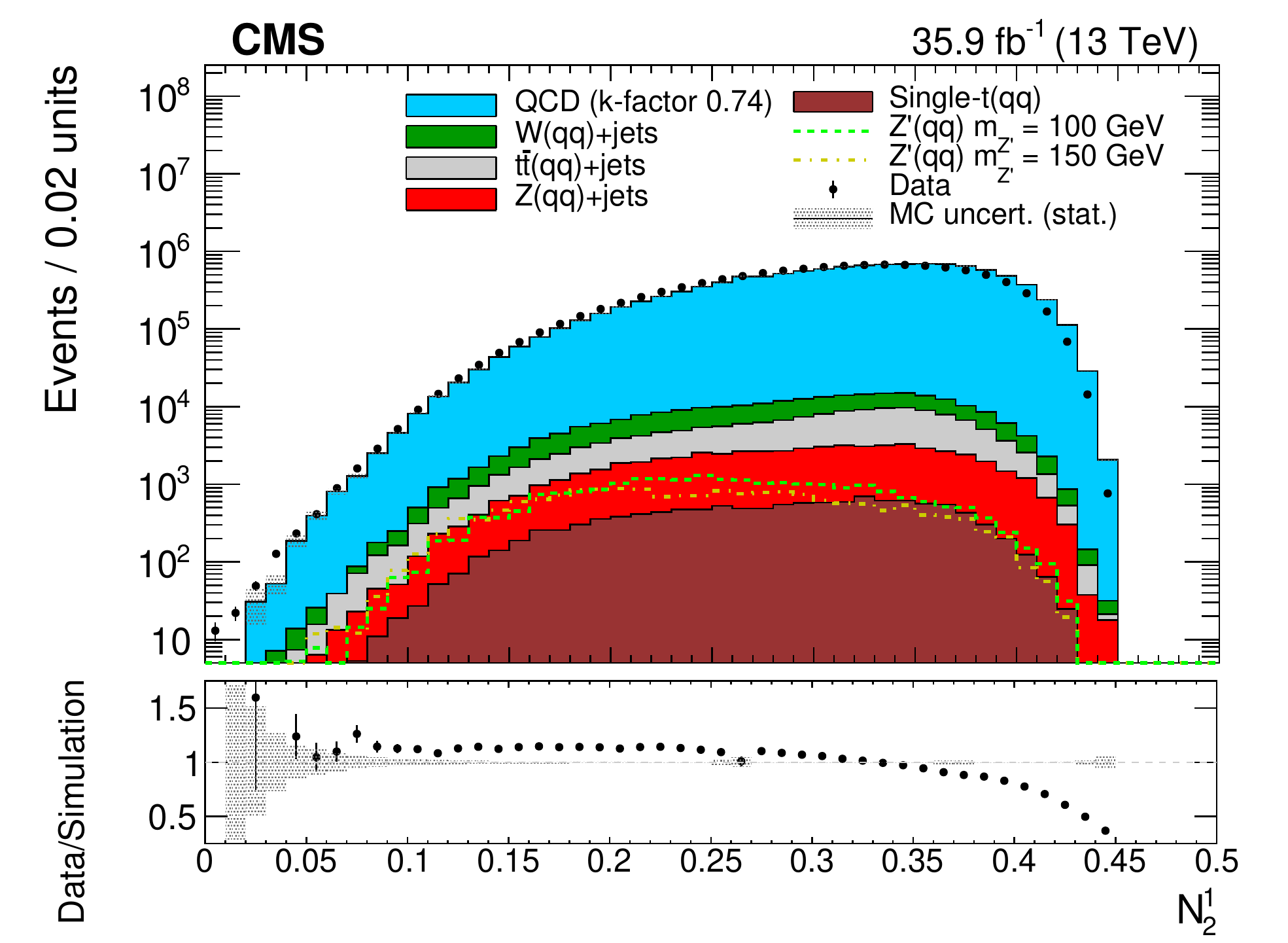} \hfil
\includegraphics[width=0.45\textwidth]{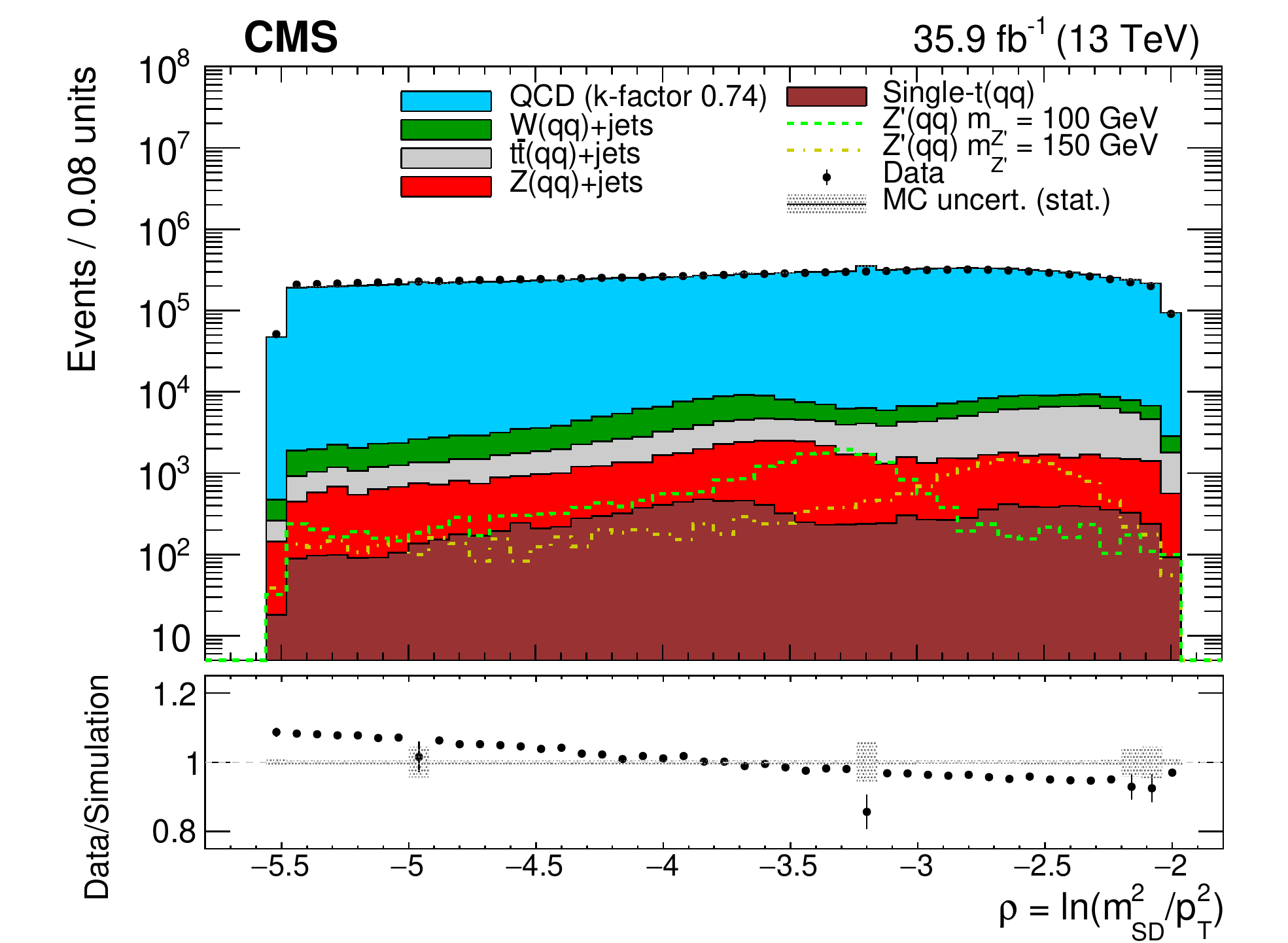}\\
\includegraphics[width=0.45\textwidth]{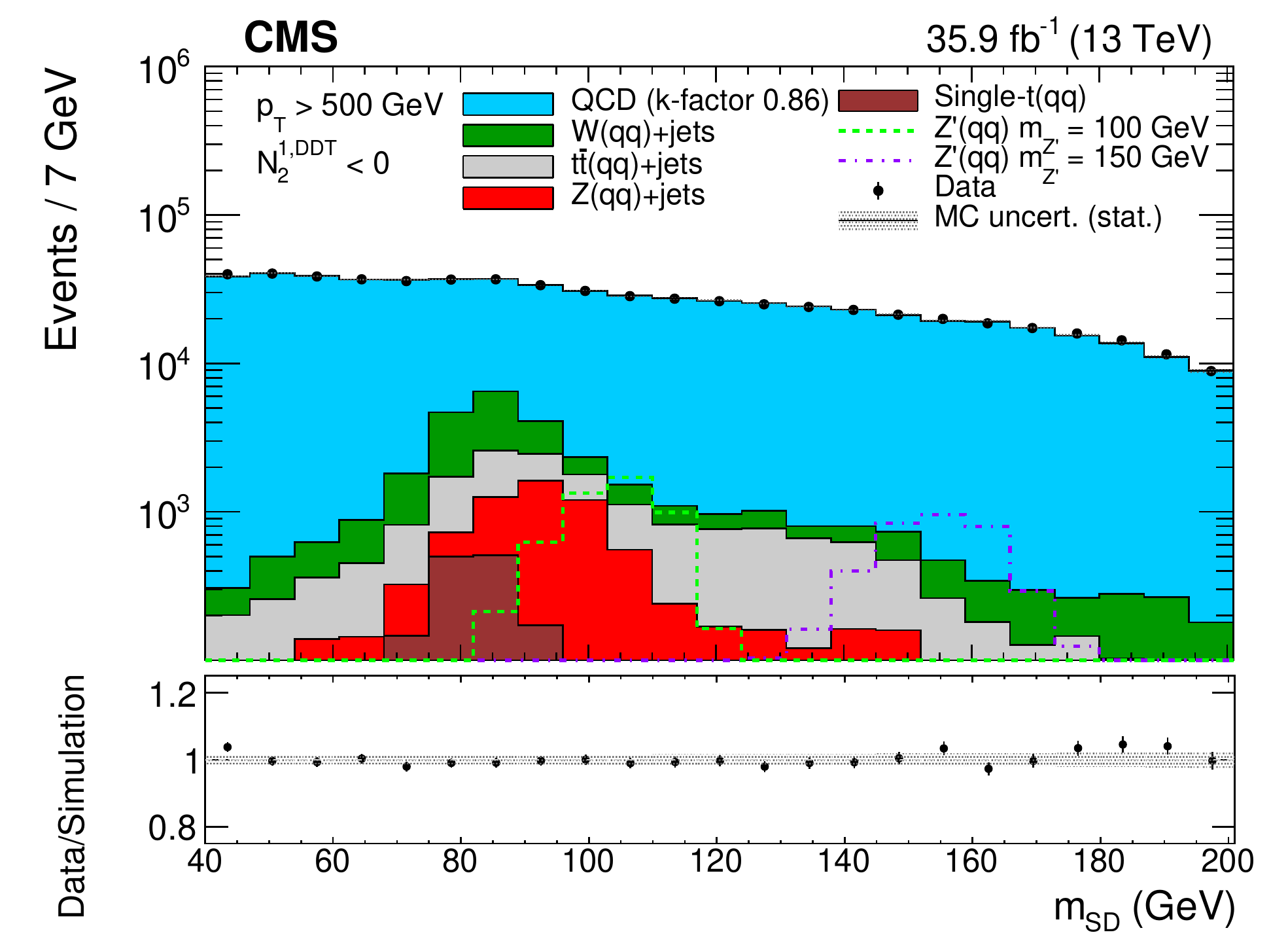} \hfil
\includegraphics[width=0.45\textwidth]{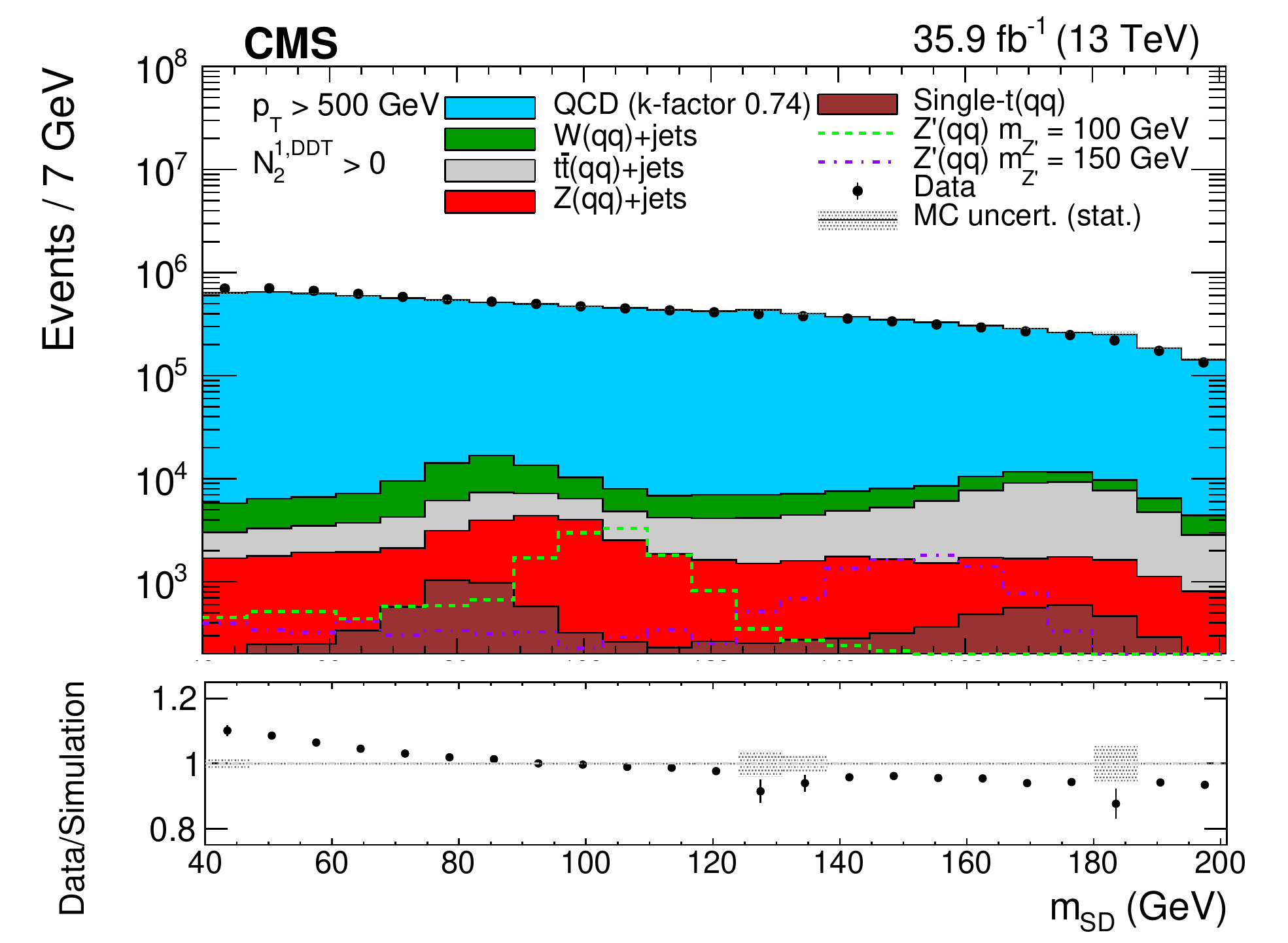}
\end{center}
\caption{
Distributions of data (points) and simulated backgrounds (histograms) 
of the leading \pt jet $N_2^{1}$ (top left) and $\rho$ (top right) observables, 
after the kinematic selection. The soft-drop jet mass distributions 
for the passing (bottom left) and failing (bottom right) region, defined by the $N_2^{1,\text{DDT}}$ selection, are also shown. 
The decorrelation ensures that the shape of the multijet mass distribution in both regions is unaffected by the 
$N_2^{1,\text{DDT}}$ selection for different \pt ranges.
Dashed lines illustrate the signal contribution for different \PZpr boson masses.
The multijet processes (QCD) dominate the background component, with subdominant contributions from inclusive SM \PW,~\PZ,
and \ttbar and single top quark processes.
The QCD simulation is scaled by an overall factor of 0.74 to match the data yield.
Residual differences between data and simulation 
demonstrate the need for a background estimation method based on control samples in data.
\label{fig:aux_stack}
}
\end{figure}

\begin{figure}[!htb]
\begin{center}
\includegraphics[width=0.75\textwidth]{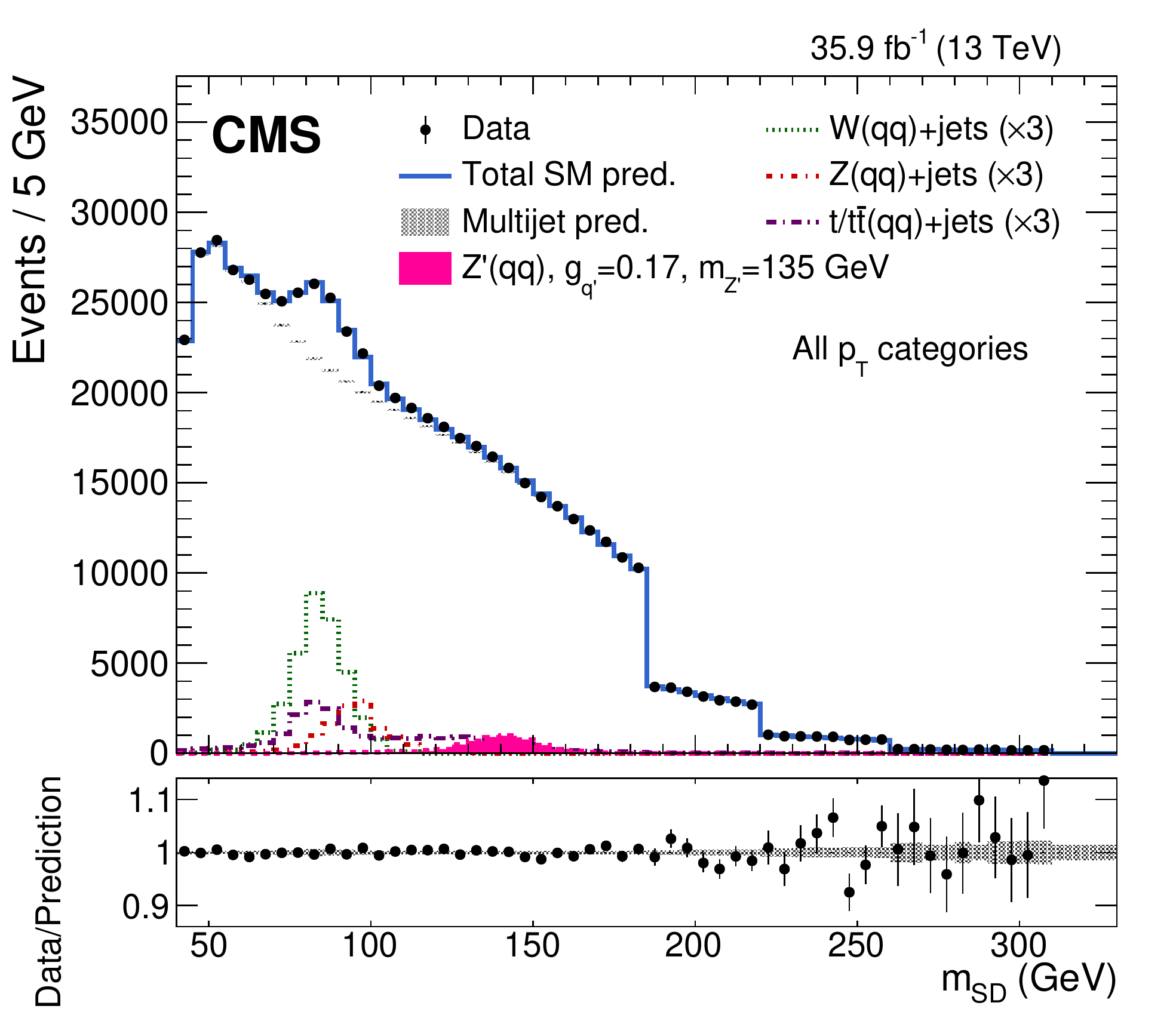}
\end{center}
\caption{
Soft-drop jet mass distribution for the passing region and combined \pt categories.
The multijet background prediction in the passing region is obtained using the failing region and the pass--fail ratio $R_{\rm p/f}(m_\text{SD},\pt)$.
Data are shown as black points.
The multijet background prediction, including uncertainties, is shown by the shaded bands.
Contributions from the \PW~and~\PZ~boson, and top quark background processes 
are shown, scaled up by a factor of 3 for clarity.
A hypothetical \PZpr boson signal at a mass of 135\GeV is also indicated.
The features at 45, 185, 220 and 255\GeV in the $m_{\rm SD}$ distribution are due to the kinematic selection on $\rho$, which affects each \pt category differently.
In the bottom panel, the ratio of the data to the background prediction, including uncertainties, is shown.
\label{fig:aux_postfit_s_135_cat6}}
\end{figure}

\begin{figure}[hbtp]
\begin{center}
\includegraphics[width=0.75\textwidth]{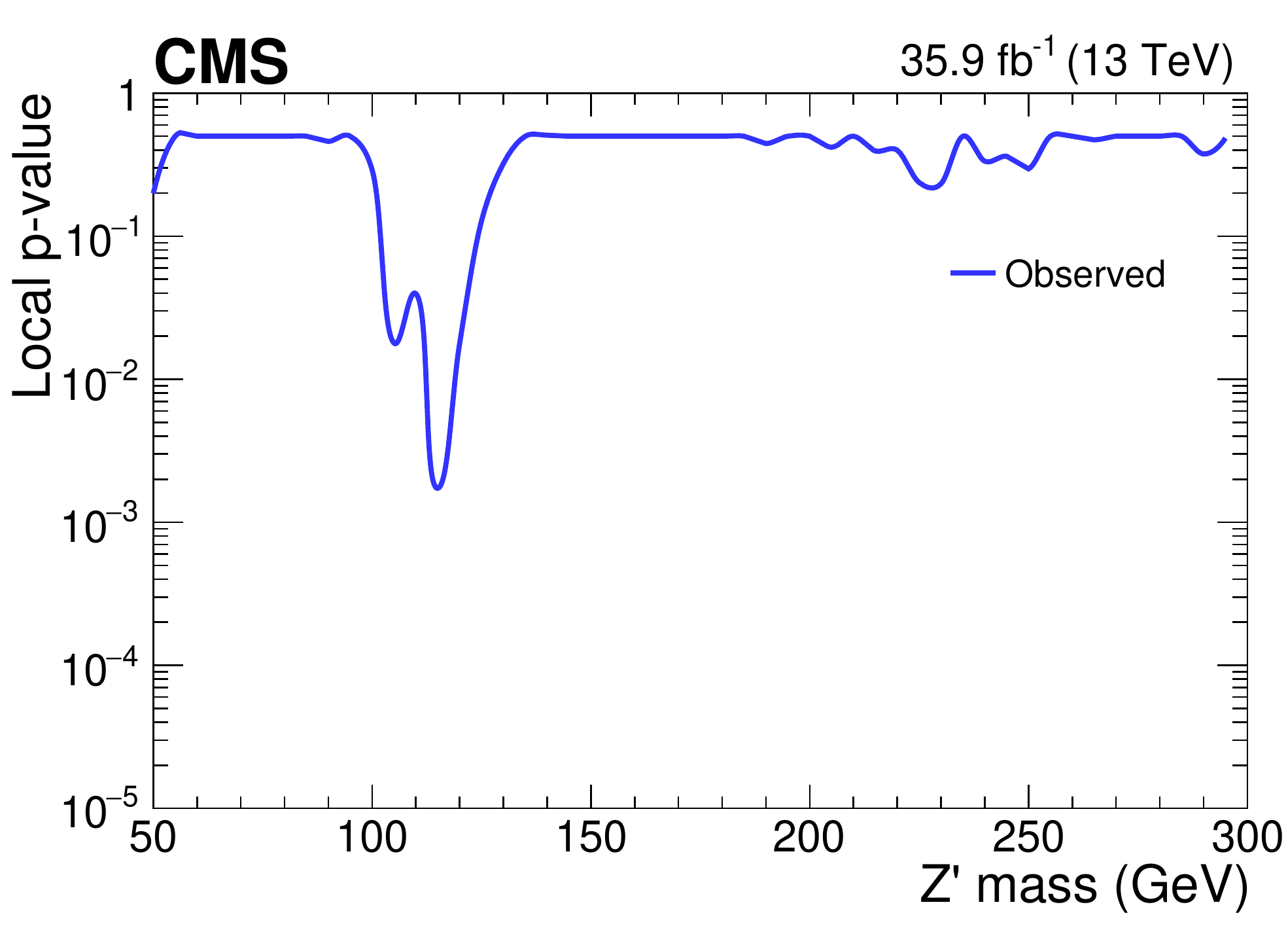}
\caption{
The observed $\text{p}$-value, obtained from the fit to data, as a function of the \PZpr boson mass. 
The maximum local observed $\text{p}$-value, at 115\GeV, is $1.72\times 10^{-3}$ and 
corresponds to $2.9$ standard deviations from the background-only expectation, 
and the global $\text{p}$-value, calculated over the probed mass range, corresponds to $0.0138$ and $2.2$ standard deviations.
\label{fig:aux_pvalue}
}
\end{center}
\end{figure}

\cleardoublepage \section{The CMS Collaboration \label{app:collab}}\begin{sloppypar}\hyphenpenalty=5000\widowpenalty=500\clubpenalty=5000\input{EXO-17-001-authorlist.tex}\end{sloppypar}
\end{document}

%% file: EXO-17-001-authorlist.tex
\textbf{Yerevan Physics Institute,  Yerevan,  Armenia}\\*[0pt]
A.M.~Sirunyan, A.~Tumasyan
\vskip\cmsinstskip
\textbf{Institut f\"{u}r Hochenergiephysik,  Wien,  Austria}\\*[0pt]
W.~Adam, F.~Ambrogi, E.~Asilar, T.~Bergauer, J.~Brandstetter, E.~Brondolin, M.~Dragicevic, J.~Er\"{o}, M.~Flechl, M.~Friedl, R.~Fr\"{u}hwirth\cmsAuthorMark{1}, V.M.~Ghete, J.~Grossmann, J.~Hrubec, M.~Jeitler\cmsAuthorMark{1}, A.~K\"{o}nig, N.~Krammer, I.~Kr\"{a}tschmer, D.~Liko, T.~Madlener, I.~Mikulec, E.~Pree, N.~Rad, H.~Rohringer, J.~Schieck\cmsAuthorMark{1}, R.~Sch\"{o}fbeck, M.~Spanring, D.~Spitzbart, W.~Waltenberger, J.~Wittmann, C.-E.~Wulz\cmsAuthorMark{1}, M.~Zarucki
% ...
\vskip\cmsinstskip
% ...
1:~~Also at Vienna University of Technology, Vienna, Austria\\
% ...
